\documentclass{elsart}
\usepackage{epsfig}

\journal{{\tt Astroparticle Physics}}

\begin{document}

\begin{frontmatter}

\title{KASCADE measurements of energy spectra for elemental groups of cosmic
rays: Results and open problems}

\author[uni]{T. Antoni},
\author[fzk]{W.D. Apel},
\author[fzk]{A.F. Badea\thanksref{r1}}, 
\author[fzk]{K. Bekk},
\author[rum]{A. Bercuci},
\author[fzk,uni]{J. Bl\"umer},
\author[fzk]{H. Bozdog},
\author[rum]{I.M. Brancus}, 
\author[arm]{A. Chilingarian},
\author[fzk]{K. Daumiller},
\author[fzk]{P. Doll},
\author[fzk]{R. Engel}, 
\author[fzk]{J. Engler}, 
\author[fzk]{F. Fe{\ss}ler}, 
\author[fzk]{H.J. Gils},
\author[uni]{R. Glasstetter\thanksref{r2}}, 
\author[fzk]{A. Haungs},
\author[fzk]{D. Heck},
\author[uni]{J.R. H\"orandel}, 
\author[uni,fzk]{K.-H. Kampert\thanksref{r2}}, 
\author[fzk]{H.O. Klages},
\author[fzk]{G. Maier\thanksref{r3}},
\author[fzk]{H.J. Mathes}, 
\author[fzk]{H.J. Mayer},
\author[fzk]{J. Milke},
\author[fzk]{M. M\"uller}, 
\author[fzk]{R. Obenland},
\author[fzk]{J. Oehlschl\"ager}, 
\author[fzk]{S. Ostapchenko\thanksref{r4}}, 
\author[rum]{M. Petcu},
\author[fzk]{H. Rebel},
\author[pol]{A. Risse},
\author[fzk]{M. Risse},
\author[uni]{M. Roth},
\author[fzk]{G. Schatz},
\author[fzk]{H. Schieler}, 
\author[fzk]{J. Scholz},
\author[fzk]{T. Thouw},
\author[fzk]{H. Ulrich\corauthref{corr}},
\author[fzk]{J. van Buren},
\author[arm]{A. Vardanyan},
\author[fzk]{A. Weindl},
\author[fzk]{J. Wochele}, 
\author[pol]{J. Zabierowski}

\address[uni]{Institut f\"ur Experimentelle Kernphysik, Universit\"at 
Karlsruhe, 76021~Karlsruhe, Germany}
\address[fzk]{Institut f\"ur Kernphysik, Forschungszentrum Karlsruhe,
76021~Karlsruhe, Germany}
\address[rum]{National Institute of Physics and Nuclear Engineering, 
7690~Bucharest, Romania}
\address[arm]{Cosmic Ray Division, Yerevan Physics Institute, 
Yerevan~36, Armenia}
\address[pol]{Soltan Institute for Nuclear Studies, 90950~Lodz, Poland}

\corauth[corr]{corresponding author, {\it E-mail address:}
Holger.Ulrich@ik.fzk.de}
\thanks[r1]{on leave of absence from Nat.\ Inst.\ of Phys.\ and 
Nucl.\ Engineering, Bucharest, Romania}
\thanks[r2]{now at Fachbereich Physik, Universit\"at Wuppertal, 
42097~Wuppertal, Germany}
\thanks[r3]{now at University of Leeds, LS2 9JT Leeds, United Kingdom}
\thanks[r4]{on leave of absence from Moscow State University, 
119899~Moscow, Russia} 

\ifx AA
\makeatletter
\begingroup
  \global\newcount\c@sv@footnote
  \global\c@sv@footnote=\c@footnote     
  \output@glob@notes  
  \global\c@footnote=\c@sv@footnote     
  \global\t@glob@notes={}
\endgroup
\makeatother
\fi

\newpage

\begin{abstract}
A composition analysis of KASCADE air shower data is performed by means of
unfolding the two-dimensional frequency spectrum of electron and muon numbers.
Aim of the
analysis is the determination of energy spectra for elemental groups
representing the chemical composition of primary cosmic rays. Since such an
analysis depends crucially on simulations of air showers the two
different hadronic interaction models QGSJet and SIBYLL are used for their
generation. The resulting primary energy spectra show
that the knee in the all particle spectrum is due to a steepening of the
spectra of light elements but, also, that neither of the two simulation
sets is able to describe the measured data consistently over the whole
energy range with discrepancies appearing in different energy regions.
\end{abstract}

\end{frontmatter}

\section{Introduction}
The energy spectrum of primary cosmic rays, extending over more than 12
decades in energy, follows, over a large range, a simple power law
$dJ/dE$ $\propto E^{\gamma}$ indicating its non-thermal character.
However, in the region between 1~PeV and 10~PeV
a change of the spectral index from $\gamma\approx-2.65$ to
$\gamma\approx-3.1$ occurs,
the so-called {\it knee} in the spectrum of cosmic rays. Since its
discovery \cite{knee_disc} nearly 50 years ago many measurements have
been performed in this energy range (see e.g. \cite{swordy} for
recent measurement results), but the origin of the knee
feature is still not convincingly explained.

Proposals for its origin range from astrophysical scenarios like the
change of acceleration mechanisms \cite{knee-berez,knee-stanev,knee-kobay}
at the sources of cosmic rays (supernova remnants, pulsars, etc.) or effects
due to the propagation \cite{knee-ptus,knee-cand}
inside the Galaxy (diffusion, drift, escape from the Galaxy) to particle
physics models like the interaction with relic neutrinos \cite{knee-relic}
during transport or new processes in the atmosphere
\cite{knee-nikol,knee-kazan} during air shower development.
Common to all models is the prediction of a change of composition over the
knee region. Moreover, in order to distinguish between individual models,
knowledge of the energy spectra of individual elements or at least mass groups
of primary cosmic rays is desired since the different models predict
different spectral shapes. 

Because of the low fluxes of cosmic rays only indirect measurements
via the detection of extensive air showers (EAS) induced by primary cosmic
ray particles in the atmosphere are feasible at present in the energy range
close to and above 1~PeV.
Determination of spectra for individual elements or mass groups is limited
by the large intrinsic fluctuations of EAS observables.
Furthermore, any analysis of air shower data has to rely on EAS simulations
and our limited knowledge of particle physics in the energy range of
relevance. Since the primary energies of
the showers are beyond the energy range of man-made accelerators and reactions
relevant to shower development occur in the very forward direction
not accessible in collider experiments, uncertainties in the description
of hadronic interactions in shower development are unavoidable. One has,
therefore,
to rely on the use of phenomenological interaction models which
differ in their predictions in some respect strongly, making the task
of retrieving information about individual energy spectra from air shower
data even more difficult. Approaches facing these difficulties by using
statistical methods and extensive comparisons with simulations can be found
e.g. in \cite{casacomp,eastopcomp}.

In this paper we present an analysis of
the classical EAS observables, electron and muon numbers, which deals 
with these problems. Because of the high accuracy of 
the KASCADE experiment the presented method, based on unfolding
procedures, is capable of reconstructing energy spectra for five elements
representing different mass groups of primary cosmic ray particles.
The analysis is performed twice using simulations with two different
high energy hadronic interaction models, QGSJet \cite{qgsjet} and
SIBYLL \cite{sibyll}. This approach gives
also a lower limit of the uncertainties due to the modelling of
hadronic interactions.
It turns out that the analysis is sensitive
to the different models allowing to identify inconsistencies
between simulations and data and to give hints to improve the models.

After a brief description of the experimental setup and the data used
in Section \ref{kascade_data_sec} the idea and the approach of the
analysis are outlined in Section \ref{analysis_outline}. Here the main
objective is the formulation of the relation between the measured
two-dimensional shower
size spectrum and the primary energy spectra as matrix equation.
Mathematical details of this procedure are given in Appendix \ref{mathe}.
In this
equation all relevant EAS and reconstruction properties are contained in the
so-called response matrix. Section \ref{simulations} deals with the
description of the distributions necessary for the calculation of the matrix
elements. In Section \ref{strategy} unfolding as a method for solving the
matrix equation is introduced whereas in Section \ref{decexamp} its
application to Monte Carlo data is discussed. The results of the unfolding
analysis applied to measured data are presented in Section
\ref{results_sec} and discussed in Section \ref{discuss_sec},
followed by the conclusions.
\section{The KASCADE experiment and data selection}
\label{kascade_data_sec}
The KASCADE (Karlsruhe Shower Core and Array Detector) experiment 
\cite{kas} investigates 
air showers in a primary energy range from 100~TeV to 100~PeV and 
measures  a large number of observables for each event:
electrons, muons at four energy thresholds, and energy and number of hadrons.
The main detector components of KASCADE are the field array \cite{kas}, 
the central detector \cite{calor,MWPC}, and the muon tracking
detector \cite{MTD}.
In the present analysis only data from the field array is used.
Detailed descriptions of the experimental setup and reconstruction
procedures of the main shower observables can be found in
\cite{kas,kas-lat}.

\begin{figure}[t]
\centering\epsfig{file=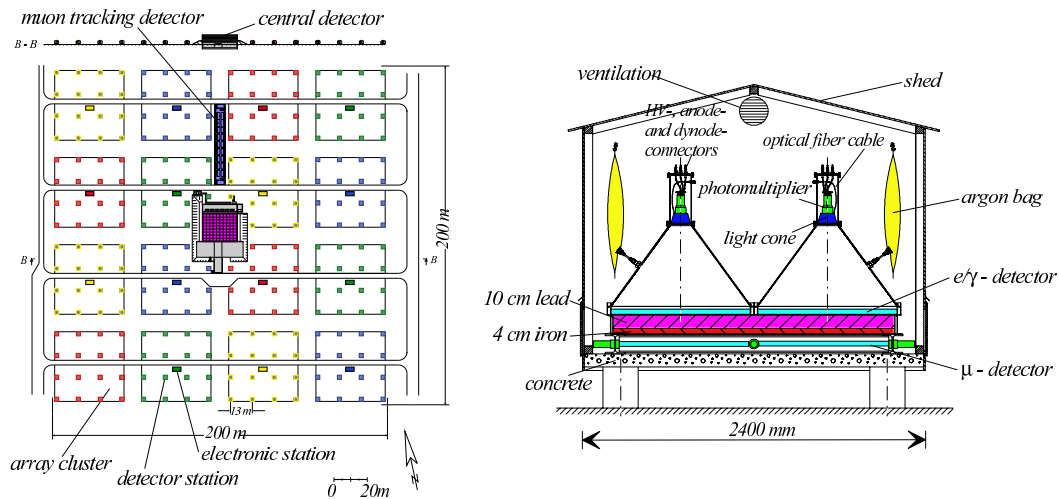,width=\linewidth}
\caption{Left: layout of the KASCADE air shower experiment. Right:
sketch of a detector station with shielded and unshielded
scintillation detectors.}
\label{KASCADE}
\end{figure}

The field array measures
electrons and muons ($E_\mu>230$~MeV)
in the shower separately using an array of 252 detector stations 
containing shielded and unshielded detectors, arranged
on a square grid of $200\times200$~m$^2$ with a spacing of 13~m. These
stations are organized in 16 so-called clusters, each consisting of 16
stations in the outer part and 15 stations in the inner part of the
array, 
respectively.
Fig.~\ref{KASCADE} displays a sketch of the
installation and of a detector station.

\begin{figure}[t]
\centering\epsfig{file=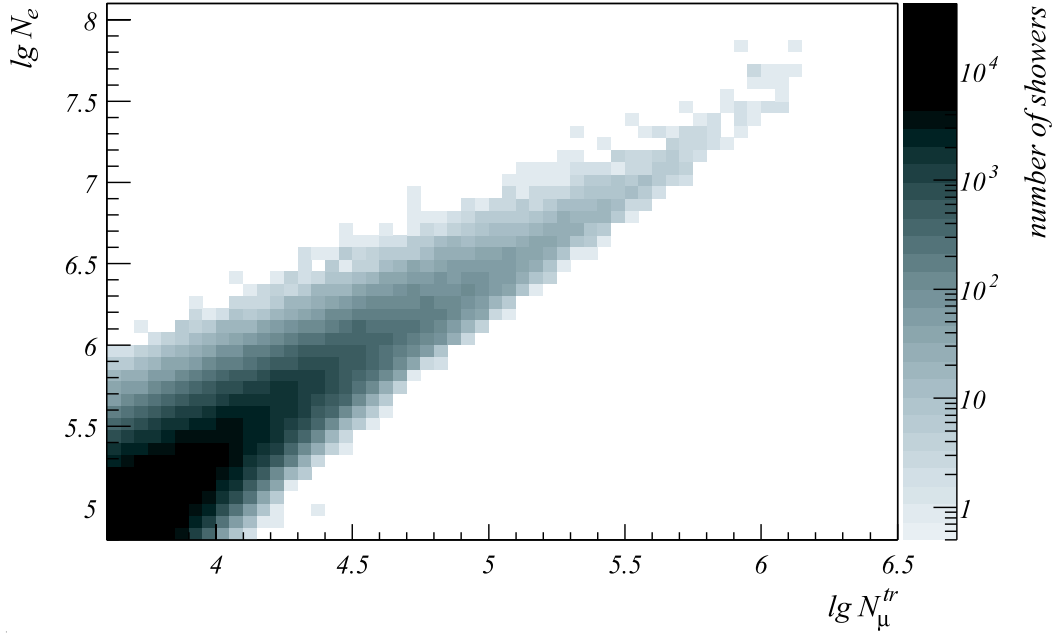,width=\linewidth}
\caption{Two-dimensional shower size spectrum used in the analysis. The
range in $\lg N_{e}$ and $\lg N_{\mu}^{tr}$ is chosen to avoid influences
of inefficiencies.}
\label{useddata_pic}
\end{figure}

The array observables used in the following are the total electron number
$N_{e}$ and the truncated muon number $N_{\mu}^{tr}$. The latter is 
the number of muons with distances to the shower core between 40~m and 200~m.
Input of the analysis is the two-dimensional shower size distribution 
with respect to these two observables. It is displayed in
Fig.~\ref{useddata_pic}.
The zenith angle of the showers in the analysis is restricted to
values between $0^{\circ}$ and $18^{\circ}$. In order to ensure a high quality
of the reconstructed shower observables the following cuts are applied:
\begin{itemize}
\item The location of the reconstructed shower core has to lie inside a circle
of 91 m radius around the center of the array. In this way an erroneous
reconstruction of showers with cores outside the array can be mostly avoided.
\item The reconstructed age-parameter $s$ of the fit with the NKG-function to
the lateral distribution of electrons has to be inside the interval
$0.2<s<2.1$. Values larger or
smaller correspond to poorly reconstructed showers which are mostly small but
may be reconstructed with large shower sizes \cite{kas-lat}.
\item Only measurement runs with all clusters active are considered. Missing
clusters during measurement strongly influence the measurement and
reconstruction thresholds.
\item
To avoid threshold effects only showers with showers sizes
$\lg N_{e}\geq4.8$ and $\lg N_{\mu}^{tr}\geq3.6$ are included.
\end{itemize}
The total number of events remaining after these cuts amounts to
$6.9\cdot10^{5}$ and
the effective measurement time adds up to 900 days. This rather small number
of remaining showers is due to the severe cuts applied in order to
guarantee a high data quality. As will be seen in the following,
the remaining statistical base is sufficient and not the limiting
factor for the reliability of the results.
\section{Outline of the analysis}
\label{analysis_outline}
Starting point of the analysis is the two-dimensional shower size spectrum
and the contents (number of events) of the histogram cells displayed in
Fig.~\ref{useddata_pic}. In the following each cell of the shower size
spectrum is labeled by a single index $i$ for identification.
The number of events in each cell $i$ results from the superposition of
contributions 
induced by different primary particles with various energies. In this sense
information about the primary energy spectra of all particle
types is present in each cell and the analyzing task is to disentangle 
this information.
\clearpage

Mathematically the content of a specific cell $i$ of the two-dimensional
spectrum, 
i.e. the number of showers $N_{i}$ with shower sizes
$(\lg N_{e},\lg N_{\mu}^{tr})_{i}$ of cell $i$, is related to the flux of
primary cosmic ray elements via the integral equation:
\begin{equation}
\hspace{-0.8cm}
N_{i} = 2\pi A_{s}T_{m}\sum_{A=1}^{N_{A}}
\int_{0^{\circ}}^{18^{\circ}}
\int_{-\infty}^{+\infty} \frac{{\mathrm{d}}J_{A}}{{\mathrm{d}}\lg E} \/
p_{A}((\lg N_{e},
\lg N_{\mu}^{tr})_{i} | \lg E) \sin\theta\cos\theta\,
{\mathrm{d}}\lg E\,{\mathrm{d}}\theta
\label{integral1}
\end{equation}
where ${\mathrm{d}}J_{A}/{\mathrm{d}}\lg E$ denotes the differential flux of
an element with mass number $A$ and the summation is carried out for all
elements present in the primary cosmic radiation. The conditional probability
$p_{A}$ describes the probability to measure a shower of primary energy
$\lg E$ and primary mass $A$ with shower sizes
$(\lg N_{e},\lg N_{\mu}^{tr})_{i}$.
Measurement time $T_{m}$ and sampling area $A_{s}$
can be treated as constants.
For the data range considered no dependence on
azimuth angle is found which results in the factor of $2\pi$.
Any dependence on solid angle is therefore reduced to the integration over
zenith angle ranging from $0^{\circ}$ to $18^{\circ}$.

The probability $p_{A}$ itself is an integral:
\begin{equation}
p_{A} = \int_{-\infty}^{+\infty}
\int_{-\infty}^{+\infty}
s_{A} \epsilon_{A} r_{A} \,d\lg N_{e}^{true} \,d\lg N_{\mu}^{tr,true}
\label{integral2}
\end{equation}
where $s_{A} = s_{A}(\lg N_{e}^{true},\lg N_{\mu}^{tr,true} | \lg E)$ are the
intrinsic shower fluctuations describing the probability for a shower with
primary mass $A$ and energy $\lg E$ to exhibit shower sizes
$\lg N_{e}^{true}$ and $\lg N_{\mu}^{tr,true}$ at observation level.
$\epsilon_{A} = \epsilon_{A}(\lg N_{e}^{true}, \lg N_{\mu}^{tr,true})$
represents the
detection and reconstruction efficiency which depends on the true
shower sizes. The probability $r_{A} = r_{A}((\lg N_{e},
\lg N_{\mu}^{tr})_{i} |$
$\lg N_{e}^{true},\lg N_{\mu}^{tr,true})$ eventually describes the
properties of the reconstruction procedure.
It accounts for the resolution of the reconstruction algorithms
and systematic effects like under- and overestimation of the shower
sizes due to the used fit functions or saturation effects of
the detectors, e.g. (see section \ref{thereconstruction} for details).
In addition,
all these quantities (especially the shower fluctuations $s_{A}$) depend in
principle on zenith angle.

Using the notation of Eqs.~(\ref{integral1}) and (\ref{integral2}) the data
histogram of Fig.~\ref{useddata_pic} is interpreted as a system of coupled
integral equations. In order to solve this set of equations for the energy
spectra ${\mathrm{d}}J_{A}/{\mathrm{d}}\lg E$ it will be
reduced to a matrix equation. The reformulation of the integral
equations as a matrix equation is straightforward and explained in detail
in Appendix \ref{mathe}. With the data vector $\vec{Y}$, whose elements $y_{i}$
are the cell contents $N_{i}$ of Fig.~\ref{useddata_pic}, i.e.
the two-dimensional shower size spectrum, and the vector of unknowns
$\vec{X}$, which represents the energy spectra of the individual primary
particle types, the problem can be written as
\begin{equation}
\vec{Y} = {\bf R} \vec{X}
\label{simplemateq}
\end{equation}
where ${\bf R}$ is the so-called {\it response} or {\it transfer} matrix.
The elements of ${\bf R}$ relate the energy spectra to the observed shower
size spectrum via the probabilities for measuring the observables
$\lg N_{e}$, $\lg N_{\mu}^{tr}$ of an air shower induced by a primary particle
with mass $A$ and energy $\lg E$.
All physical information about the air showers
as well as the detection and reconstruction properties are
contained in the response matrix ${\bf R}$. Any results regarding energy
spectra depend crucially on the knowledge of the matrix elements. The
calculation of the matrix elements, i.e. the determination of the quantities
$s_{A}$, $r_{A}$, and $\epsilon_{A}$ was based on Monte Carlo simulations
using the CORSIKA \cite{corsika}
program.
Due to the impossibility to account in the analysis for all elements
present in cosmic rays, we confine ourselves to five
elements representing individual mass groups: hydrogen (proton), helium,
carbon (CNO-group), silicon (intermediate elements) and iron (heavy component).
\section{Determination of the matrix elements}\label{simulations}
\subsection{Simulation strategy}
For the calculation of the matrix elements one has to rely on simulations
in order to determine the shower fluctuations, efficiencies, and reconstruction
properties. The corresponding simulated distributions are parameterized to
simplify the numerical integrations.
This approach allows also the
investigation of the influence of unknown
tails of the shower fluctuations, which are poorly determined
statistically, on the result.
This gives at least an estimate of this systematic uncertainty.
The following simulation strategy is pursued:
\begin{enumerate}
\item The relevant shower size distributions are
determined and parametrized for a set of fixed primary energies. These
simulations are carried out using the CORSIKA code with the
high energy interaction models QGSJet 01
and SIBYLL 2.1. For the low energy interactions
the GHEISHA 2002 \cite{gheisha} code is used.
The electromagnetic part of the showers is simulated using the EGS4
\cite{egs4} code.
In addition the thinning
option \cite{thin} for a faster simulation
was enabled. The energy dependence of the relevant parameters of the shower
size distribution is interpolated. The simulated energies are 0.1~PeV,
0.5~PeV, 1~PeV, 3.16~PeV, 10~PeV, 31.6~PeV, 100~PeV, 316~PeV and 1~EeV and the
value of the thinning level is $\varepsilon = 10^{-6}$ for all energies.
The number of simulated showers for the corresponding energies is 8000,
6000, 4000, 3000, 2000, 1500, 1000, 750, and 500, respectively, distributed
between $0^{\circ}$ and $18^{\circ}$.
A comparison between showers simulated with different thinning levels
$\varepsilon$ and without thinning was
carried out at a primary energy of 1~PeV in order to chose a thinning level
for which inescapable additional artificial fluctuations are sufficiently
small.
The relevant shower size distributions $s_{A}^{\varepsilon}$ of the simulation
sets with thinning were tested for compatibility with the corresponding
distribution $s_{A}$ defined by the simulation set without thinning. This
was done by means of a Kolmogorov-Smirnov test.
In addition, a comparison between the shower size distributions of simulation
sets using different thinning levels was performed at primary energy 1~EeV
to cross-check the energy independence of $\varepsilon$.
\item
For the determination of efficiency and reconstruction properties a
second set of CORSIKA simulations was used which consists of 
unthinned air showers,
followed by a detailed GEANT \cite{geant} simulation of the KASCADE
detectors 
and reconstruction by the standard KASCADE reconstruction software. The
initial air showers are generated according to a continuous energy spectrum
between $10^{14}$~eV and $10^{18}$~eV following a power law with differential
index $\gamma = -2$. This procedure was also performed for the
two interaction models QGSJet 01 and SIBYLL 2.1.
\end{enumerate}
\subsection{Determination of shower fluctuations $s_{A}$}
The most important distribution for the calculation of the matrix elements
is the correlated $\lg N_{e} - \lg N_{\mu}^{tr}$ - probability distribution,
i.e. the shower fluctuations $s_{A}$. Their parameterization is carried out in
two steps. For the parameterization of the
$\lg N_{e}$ - distribution for a fixed primary particle and energy the
following function is used:
\begin{equation}
\hspace{-8mm}
p(\lg N_{e}|\lg E) = p_{0}\cdot\mathrm{erf}
\left(\displaystyle\frac{\lg N_{e}-p_{1}}
{p_{2}}\right)\cdot\mathrm{exp}\left(p_{3}\cdot(\lg N_{e}-p_{4})
\right)\cdot(p_{4}-\lg N_{e})^{p_{5}}
\label{eleprobab}
\end{equation}
Here $p(\lg N_{e}|\lg E)$ is the probability density for
$\lg N_{e}$ and $p_{0}$ is
a normalization constant. The
notation $\mathrm{erf}(x)$ is the integral of a Gaussian with mean
0 and variance 0.5 between $-\infty$ and $x$. It turned out that the parameters
$p_{3}$ and $p_{5}$ can be treated as energy independent whereas $p_{1}$,
$p_{2}$, and $p_{4}$ depend on primary energy. For values of $\lg N_{e}$
larger than $p_{4}$ the probability density is assumed to be zero.

To describe the correlation between electron number and truncated
muon number it is useful to look at the fraction $Q$ of
showers with muon number above some fixed value $\lg N_{\mu}^{tr,0}$ as
function of the
electron size $\lg N_{e}$. This fraction can be well described by an error
function with varying width
\begin{equation}
Q(\lg N_{e}, \lg N_{\mu}^{tr,0}|\lg E) =
\mathrm{erf} \left(\displaystyle\frac{\lg N_{e}-\lg N_{0}}
{p_{6}-p_{7}(\lg N_{0} - \lg N_{e})}\right) \mathrm{,}
\label{fractionprobab}
\end{equation}
where $\lg N_{0}$ is a parameter depending on the value of
$\lg N_{\mu}^{tr,0}$.
For the relation between $\lg N_{0}$ and $\lg N_{\mu}^{tr,0}$ a quadratic
dependence was assumed:
\begin{equation}
\lg N_{0} = c_{0} + c_{1}\cdot\lg N_{\mu}^{tr,0} - 
c_{2}\cdot(\lg N_{\mu}^{tr,0})^{2} \mathrm{.}
\end{equation}
Using this fraction $Q$ and the probability density $p(\lg N_{e}|\lg E)$ the
correlated probability $P(\lg N_{e}, \lg N_{\mu}^{tr}|\lg E)$ for a
shower of primary energy $\lg E$ to have shower sizes $\lg N_{e}$ and
$\lg N_{\mu}^{tr}$ can be written as
\begin{eqnarray}
\lefteqn{P(\lg N_{e},\lg N_{\mu}^{tr}|\lg E) }
\nonumber \\
\lefteqn{= \left(
Q(\lg N_{e},\lg N_{\mu}^{tr}) - Q(\lg N_{e},\lg N_{\mu}^{tr} +
d\lg N_{\mu}^{tr})\right) \,p(\lg N_{e}|\lg E) \,d\lg N_{e} }
\nonumber \\
\lefteqn{= s_{A} \,d\lg N_{e}\,d\lg N_{\mu}^{tr}} 
\label{correlprobab}
\end{eqnarray}
This parameterization of the shower fluctuations $s_{A}$ for fixed particle
type and energy is used for the numerical evaluation of the matrix elements.
\begin{figure}[b]
\begin{minipage}[b]{.50\linewidth}
\centering\epsfig{file=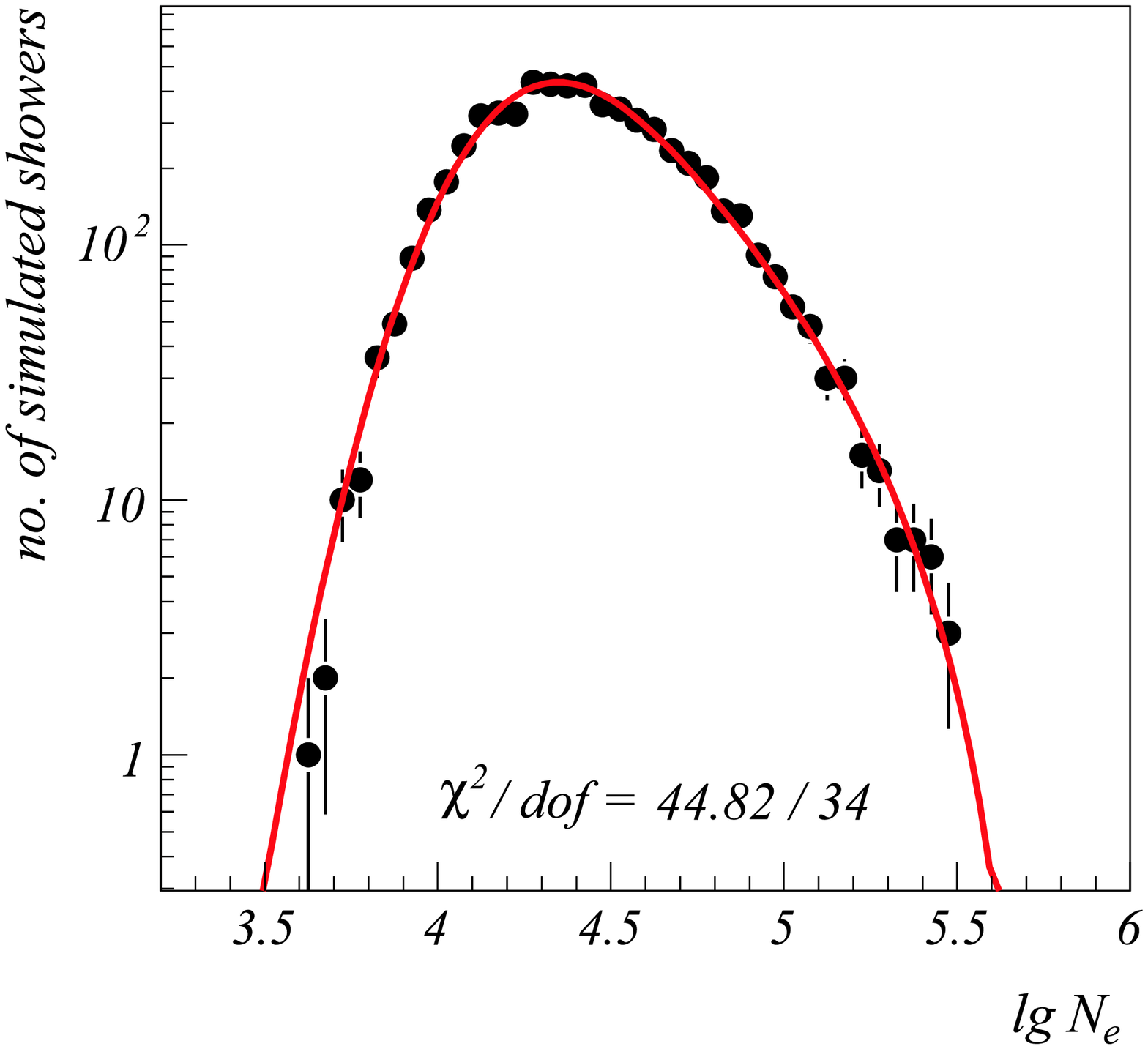,width=\linewidth}
\end{minipage}
\begin{minipage}[b]{.50\linewidth}
\centering\epsfig{file=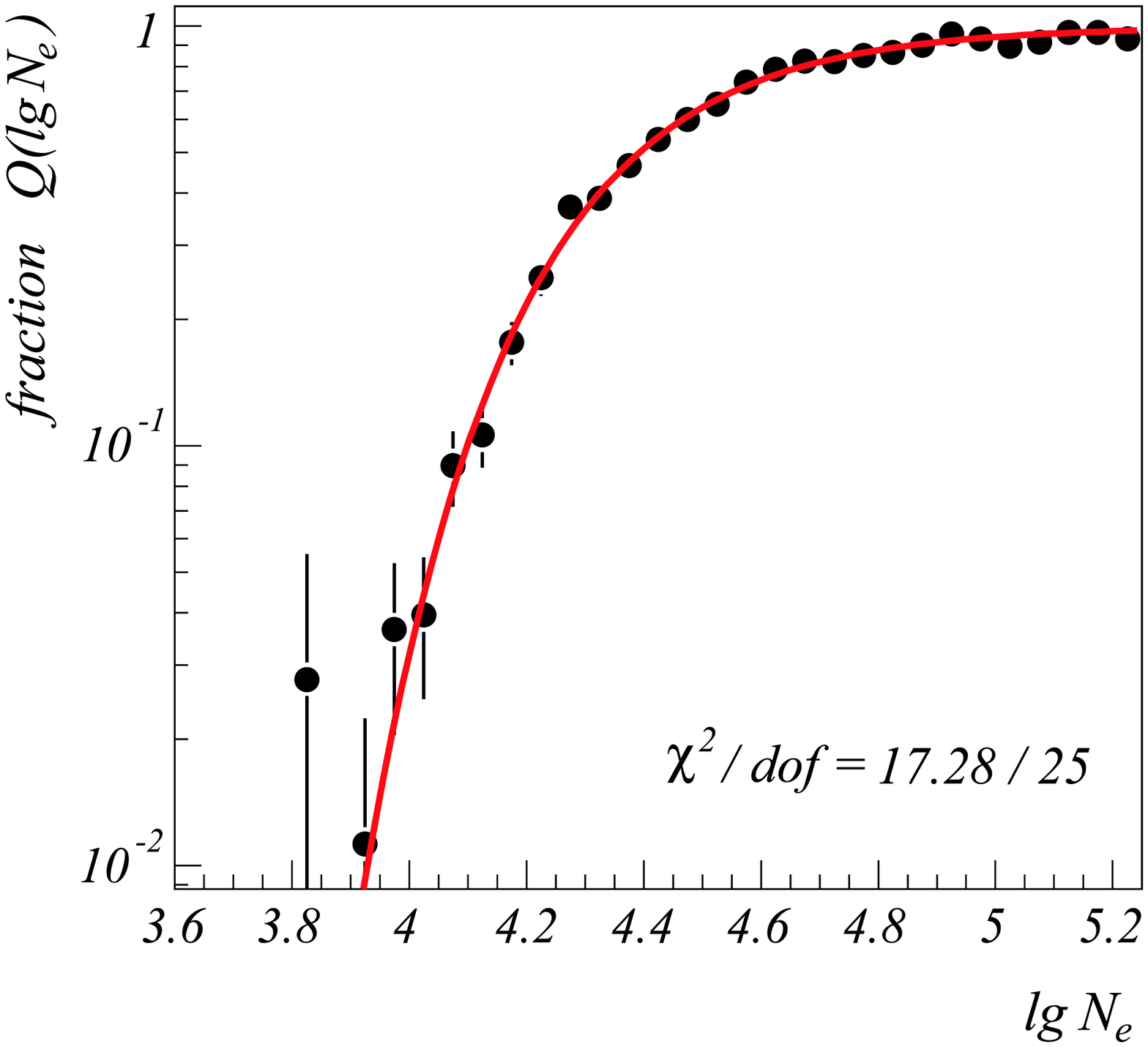,width=\linewidth}
\end{minipage}
\caption{Left: Parameterization of the electron size distribution
(proton, 0.5~PeV, QGSJet) according to Eq.~(\ref{eleprobab}).
The quality of the fit is indicated by the value of
$\chi^{2}$ per degree of freedom.
Right: Dependence of the fraction of showers, $Q$, with
$\lg N_{\mu}^{tr} > 3.2$
on $\lg N_{e}$ for the same simulated showers.
The displayed function is a fit with Eq.~(\ref{fractionprobab}).}
\label{electdistrib}
\end{figure}

To determine the free parameters of Eq.~(\ref{correlprobab}) it is
more practicable to determine first the parameters of Eq.~(\ref{eleprobab}) via
a fit to the electron size distribution and afterwards the muon parameters
of Eq.~(\ref{fractionprobab}) by fitting the truncated muon number
distribution.
The form of the latter one can be described using Eq.~(\ref{correlprobab}) by
\begin{equation}
P(\lg N_{\mu}^{tr}|\lg E) = \int_{-\infty}^{+\infty}\,\frac
{P(\lg N_{e},\lg N_{\mu}^{tr}|\lg E)}{d\lg N_{e}}\,d\lg N_{e}.
\label{muonprobab}
\end{equation}
\begin{figure}[t]
\begin{minipage}[b]{.50\linewidth}
\centering\epsfig{file=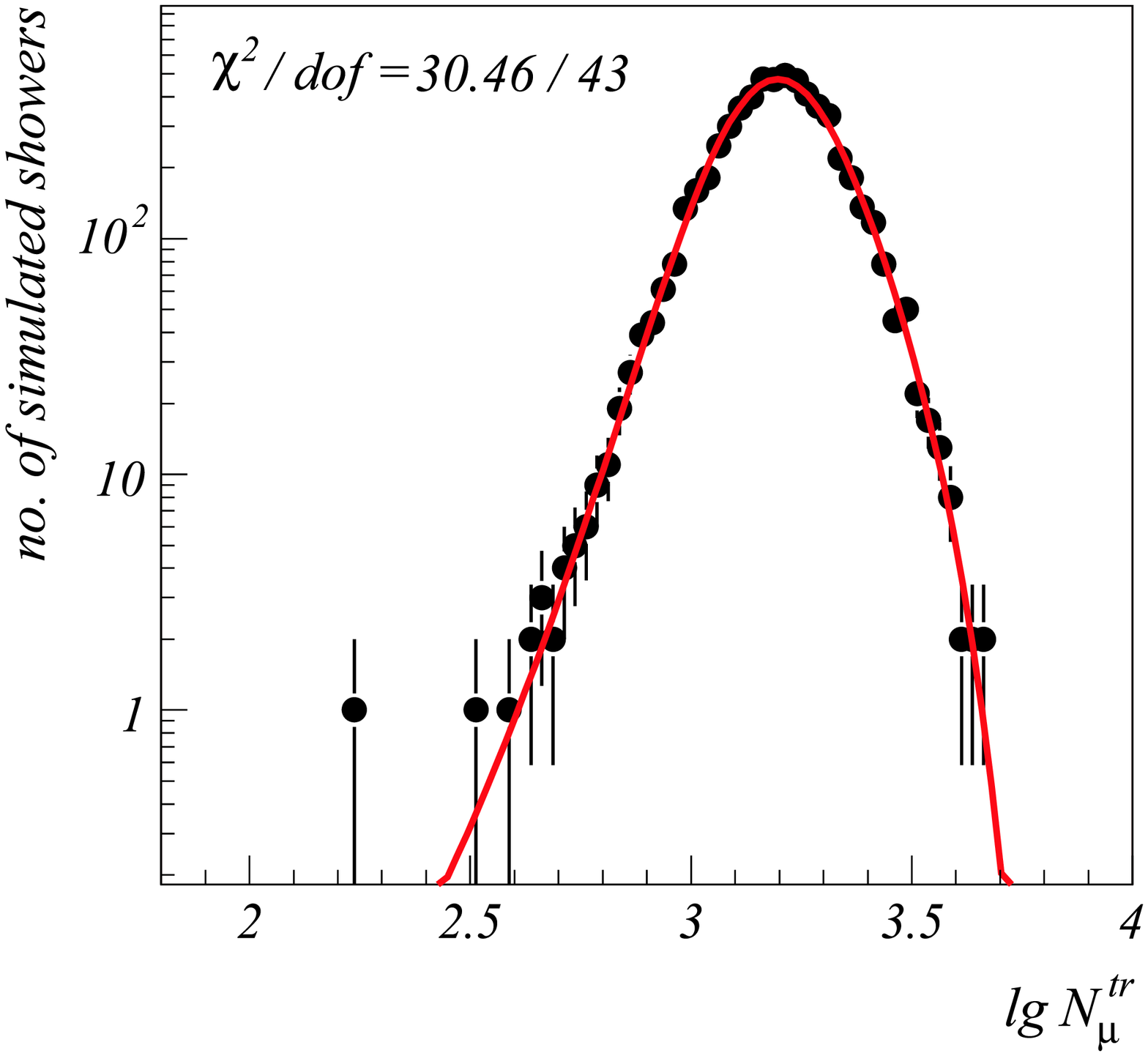,width=\linewidth}
\end{minipage}
\begin{minipage}[b]{.50\linewidth}
\centering\epsfig{file=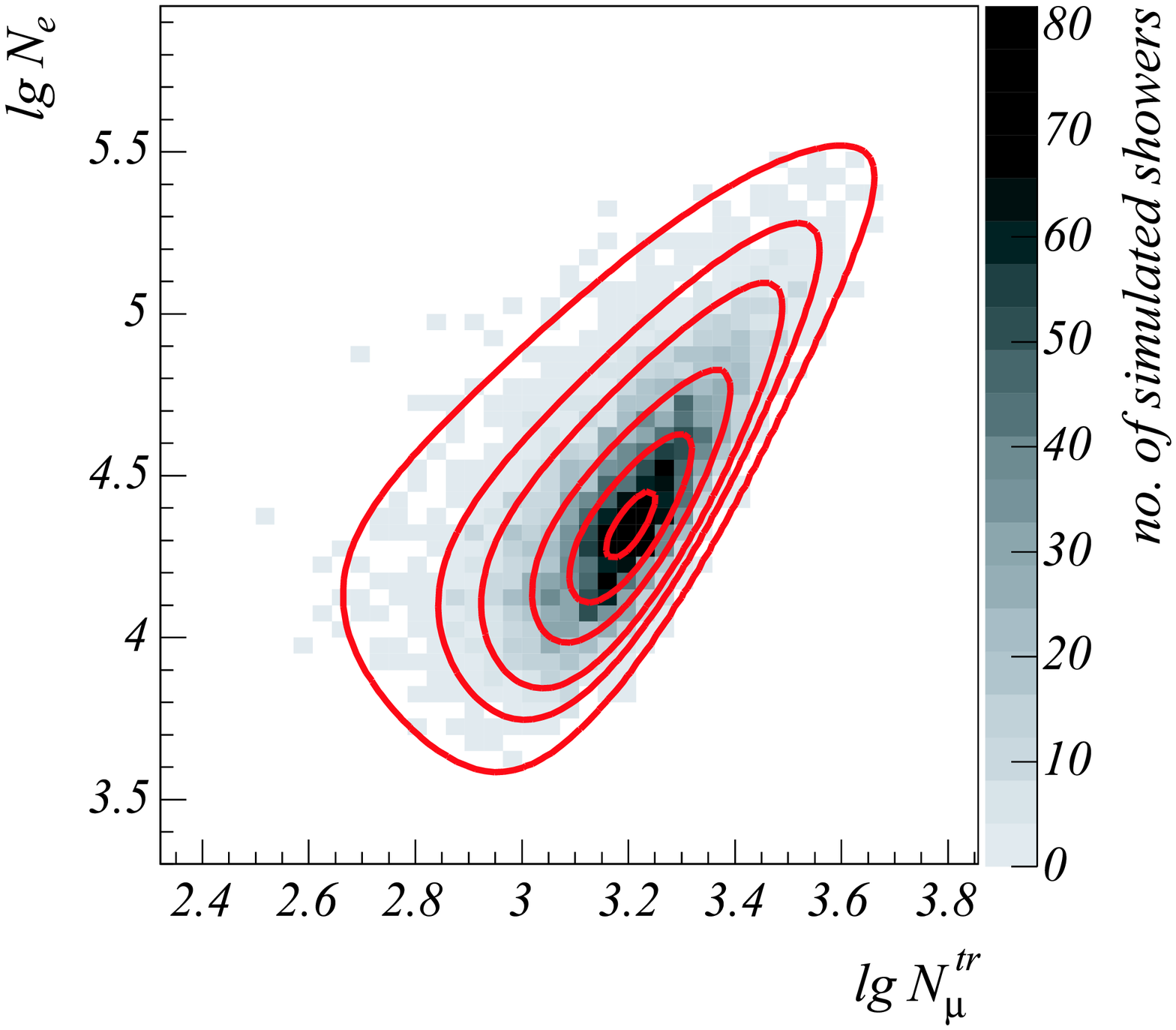,width=\linewidth}
\end{minipage}
\caption{Left: Distribution of muon number for proton induced showers
(0.5~PeV, QGSJet), together with a fit according to Eq.~(\ref{muonprobab}).
Right:
$\lg N_{e}-\lg N_{\mu}^{tr}$ distribution of simulated showers (proton,
0.5~PeV, QGSJet) and the corresponding parameterization.}
\label{muondistrib}
\end{figure}
As examples, some distributions for fixed primary energy and the corresponding
fits are illustrated in Fig.~\ref{electdistrib} and Fig.~\ref{muondistrib}.

\begin{figure}[t]
\begin{minipage}[b]{.50\linewidth}
\centering\epsfig{file=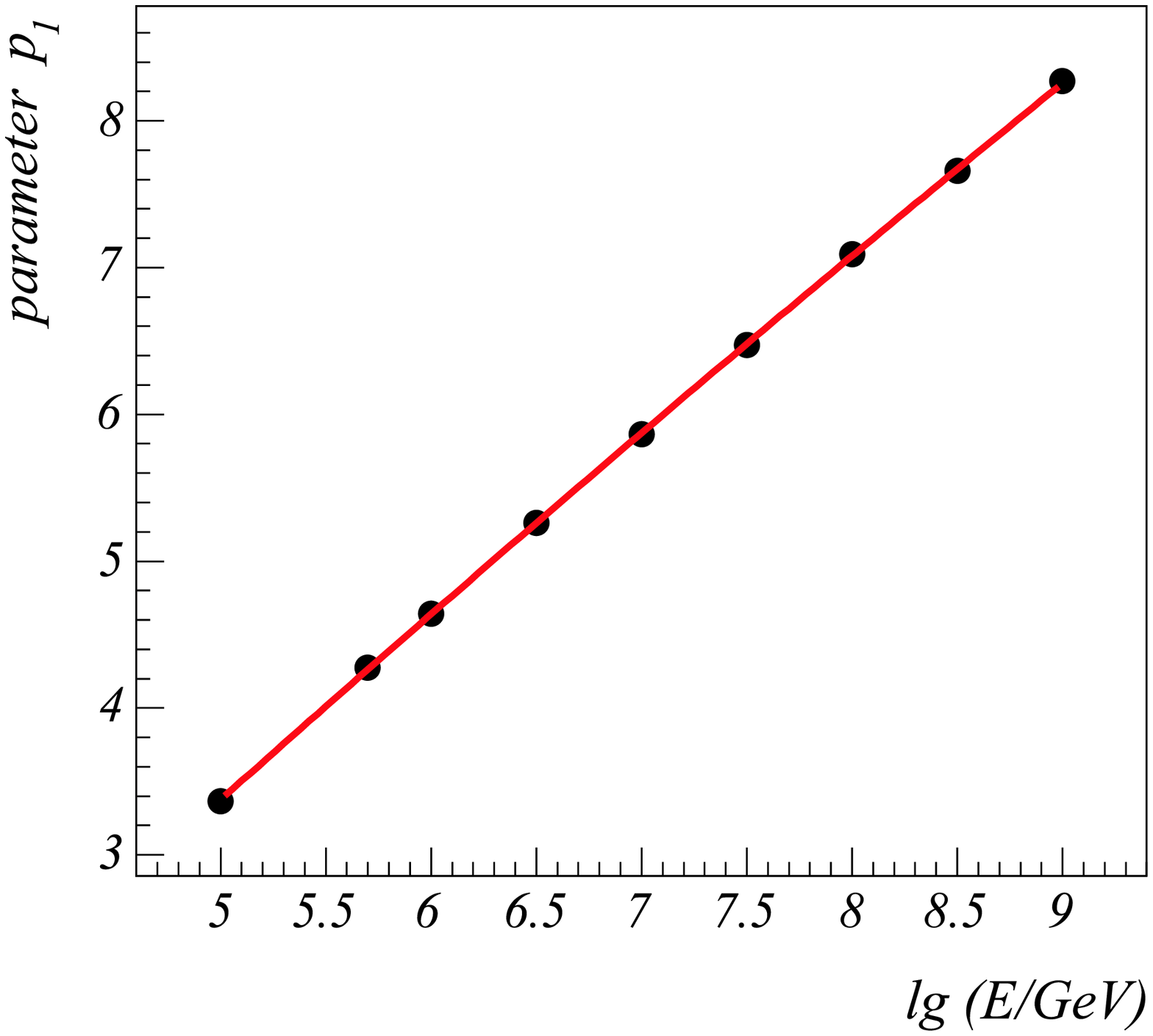,width=\linewidth}
\end{minipage}
\begin{minipage}[b]{.50\linewidth}
\centering\epsfig{file=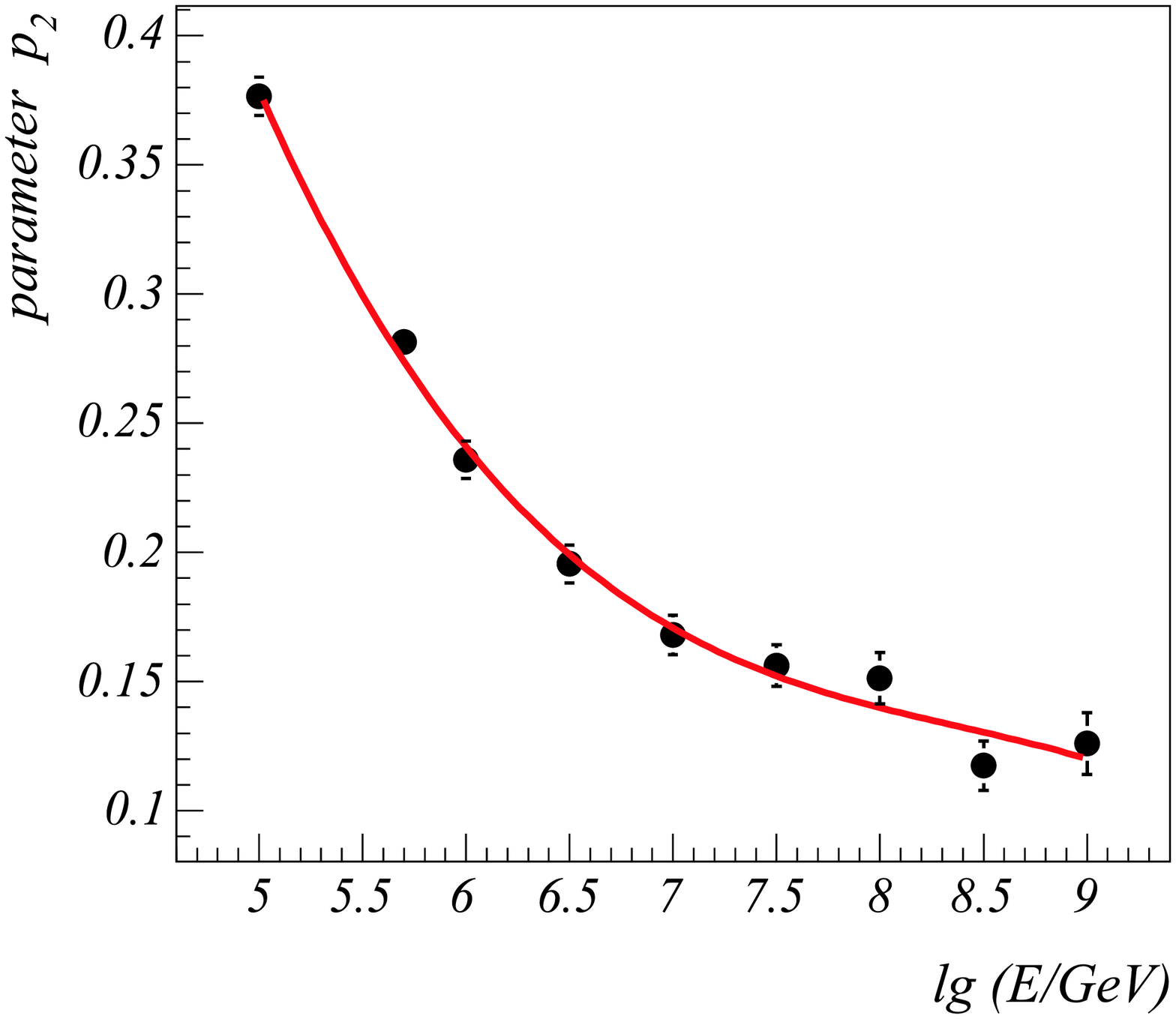,width=\linewidth}
\end{minipage}
\caption{Interpolated energy dependence of parameters $p_{1}$ and $p_{2}$ of
Eq.~(\ref{eleprobab}) for the case of proton induced showers (QGSJet
simulations).}
\label{p1p2energy}
\end{figure}

Investigation of the energy dependence of the various parameters showed
that for each primary species the parameters $p_{3}$, $p_{5}$, $p_{7}$,
$c_{1}$, and $c_{2}$ can be treated as energy independent. Furthermore,
for each
primary particle type the same values of $p_{4}$ can be used which
indicates that the electron number at shower maximum is almost 
independent of primary mass. The energy dependence of the 
remaining five parameters is interpolated using polynomials.
As an example, the parameters $p_{1}$ and $p_{2}$ for proton
induced showers (QGSJet simulations) are displayed in Fig.~\ref{p1p2energy}.
\subsection{Properties of the reconstruction}
For the investigation of the reconstruction properties, 
the set of fully simulated (no thinning) CORSIKA
showers is used, including a GEANT based simulation of the KASCADE experiment.
In these detector simulations all properties of
the detectors and the electronics are accounted for. The reliability of the
simulations was checked independently e.g. by comparison between simulated and
measured single muon spectra recorded by the array detectors. The output of
the simulations has the same data structure as measured events. Therefore,
simulated and measured showers are indistinguishable for the reconstruction
process and can be treated with the usual KASCADE
reconstruction algorithms.
\subsubsection{Estimate of $\epsilon_{A}$}
Although the data range for the following analysis is chosen in a way to
minimize influences from possible efficiency variations, it is useful 
to parametrize  the combined
trigger and reconstruction efficiencies for the calculation
of the response matrix elements. Details of trigger and reconstruction
efficiencies can be found in \cite{kas}.
Since the number of fired
detectors depends, to very good approximation, on electron shower size
only, the trigger efficiency can be well approximated by an integrated Gaussian
distribution, depending only on $\lg N_{e}^{true}$.\\
The reconstruction of a measured shower is only successful if both, electron
and muon number, can be determined.
The probability of a successful reconstruction depends on the muon number
only, as the determination of $N_{e}$ is possible for every triggered shower.
The combined efficiency $\varepsilon=\varepsilon(N_{e}, N_{\mu}^{tr})$ for
triggering the measurement and successful reconstruction can be parameterized
by the product of two error functions:
\begin{equation}
\varepsilon = {\mathrm{erf}}(a) \cdot
{\mathrm{erf}}(b) \quad {\mathrm{with}}\quad a = \frac{\lg N_{e} - p_{0}}
{p_{1}}, \quad b = \frac{\lg N_{\mu}^{tr} - p_{2}}{p_{3}}
\end{equation}
For the considered zenith angle range typical values for proton induced
showers are $p_{0} = 3.75$, $p_{1} = 0.16$, $p_{2} = 2.32$, and $p_{3} = 0.36$.
A slight dependence on primary particle type was found for the
parameters $p_{0}$ and $p_{1}$. In conclusion, all showers 
with $\lg N_{e}>4.4$ and
$\lg N_{\mu}^{tr}>3.4$ trigger the experiment and are
reconstructed successfully, regardless of primary particle type.
\subsubsection{Parameterization of $r_{A}$}\label{thereconstruction}
\begin{figure}[b]
\begin{minipage}[t]{.50\linewidth}
\centering\epsfig{file=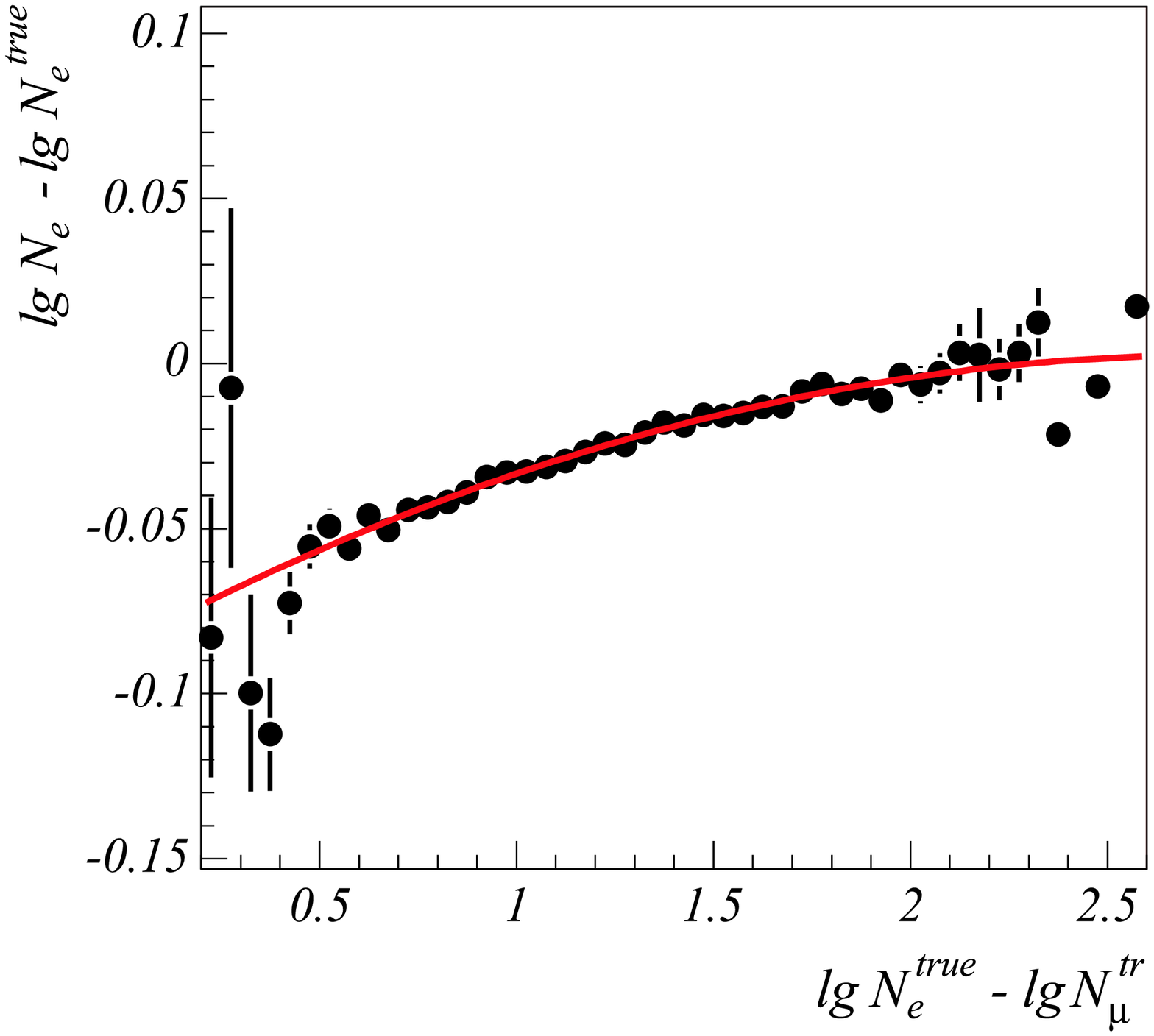,width=\linewidth}
\end{minipage}
\begin{minipage}[t]{0.50\linewidth}
\centering\epsfig{file=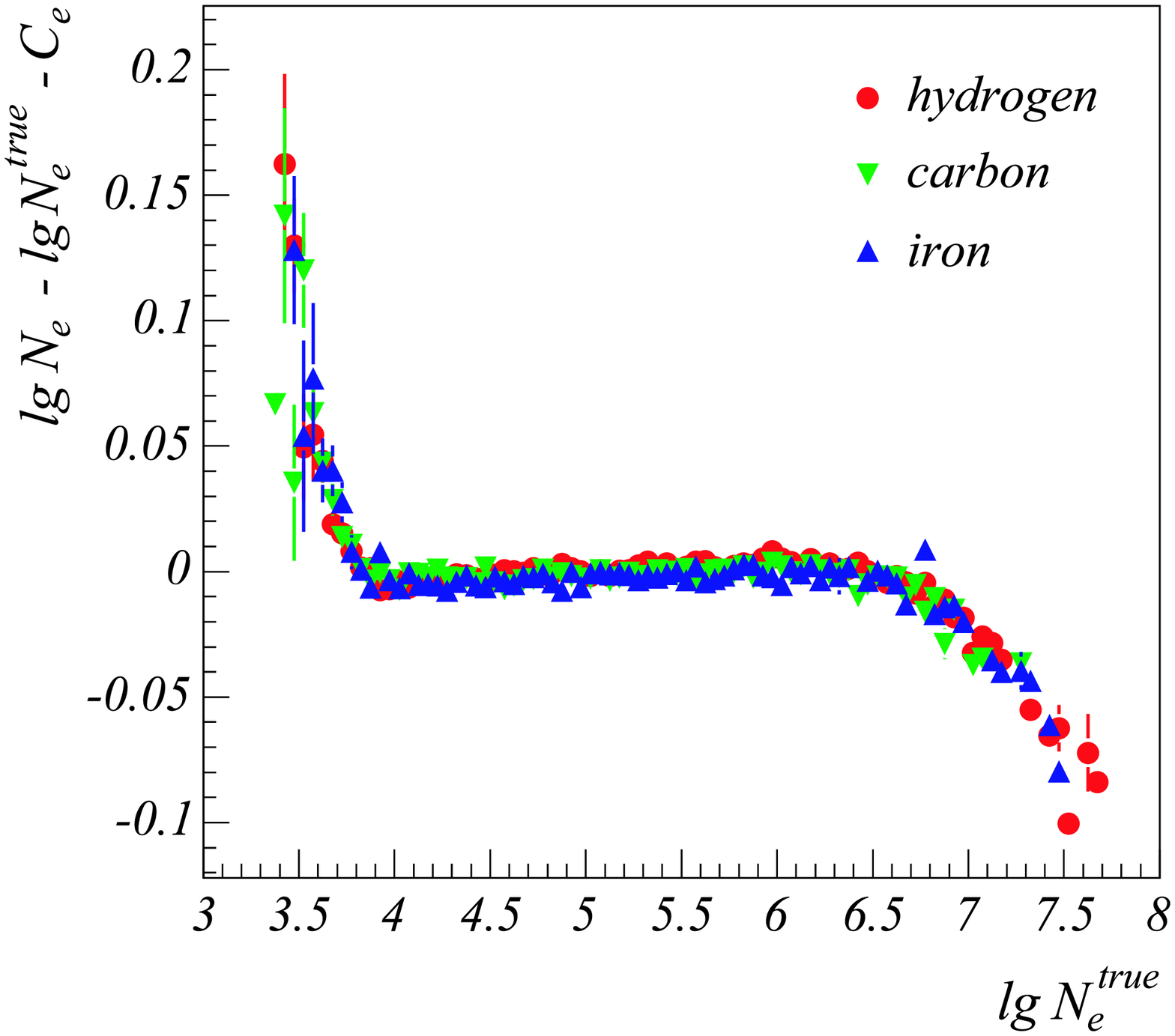,width=\linewidth}
\end{minipage}
\caption{Left: Difference between reconstructed and true electron number,
dependent on the difference $\lg N_{e}^{true} - \lg N_{\mu}^{tr}$.
Right: Remaining systematic difference after correction with the relation in
the left part of the figure. Simulations were generated using QGSJet.}
\label{elerek_syst}
\end{figure}

Of further importance for the determination of the response matrix is the
difference between reconstructed shower size and its true value.
In the case of the electron number the mean value of the difference
$\Delta\lg N_{e} = \lg N_{e}-\lg N_{e}^{true}$ shows a dependence on
the difference
$\lg N_{e}^{true}-\lg N_{\mu}^{tr}$ between true electron and
reconstructed truncated muon number. This correlation is displayed in the
left part of Fig.~\ref{elerek_syst}.
This dependence proved to be
independent of primary particle type and primary energy and is used for
a parameterization of $\Delta \lg N_{e}$ depending on
$\lg N_{e}^{true}-\lg N_{\mu}^{tr}$, which defines a correction term $C_{e}$
to be substracted from $\lg N_{e}$.
The main reasons for this systematic effect are deviations between the observed
lateral distribution and the NKG function used to determine the particle
number which has to be integrated over the whole lateral distance range.
The size of the systematic difference between true and
reconstructed electron number is strongly correlated with the shower age which
itself is strongly correlated with the ratio between electron and muon number.
A more detailed analysis of these interrelationships will be the topic of a
forthcoming paper.
\begin{figure}[t]
\begin{minipage}[t]{.50\linewidth}
\centering\epsfig{file=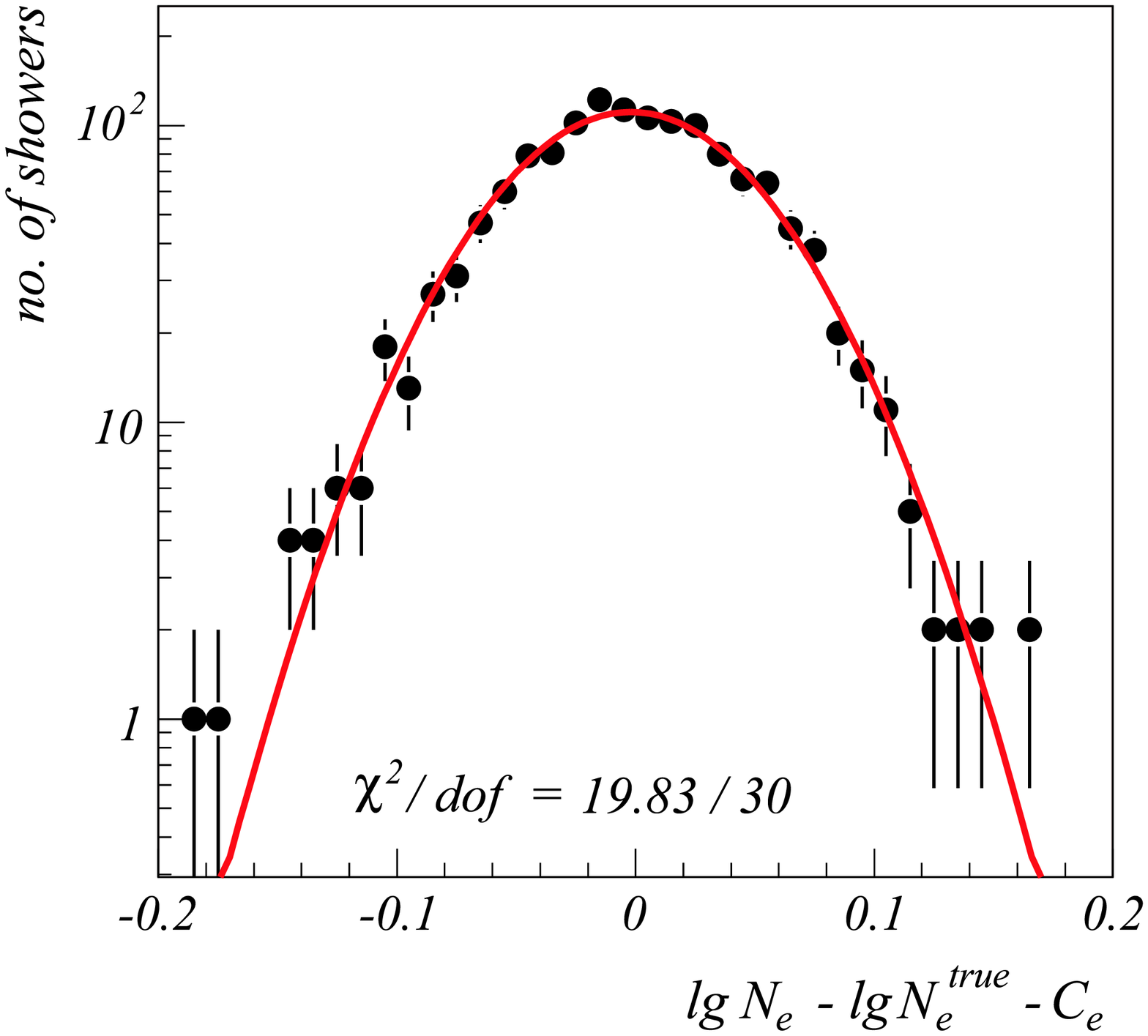,width=\linewidth}
\end{minipage}
\begin{minipage}[t]{.50\linewidth}
\centering\epsfig{file=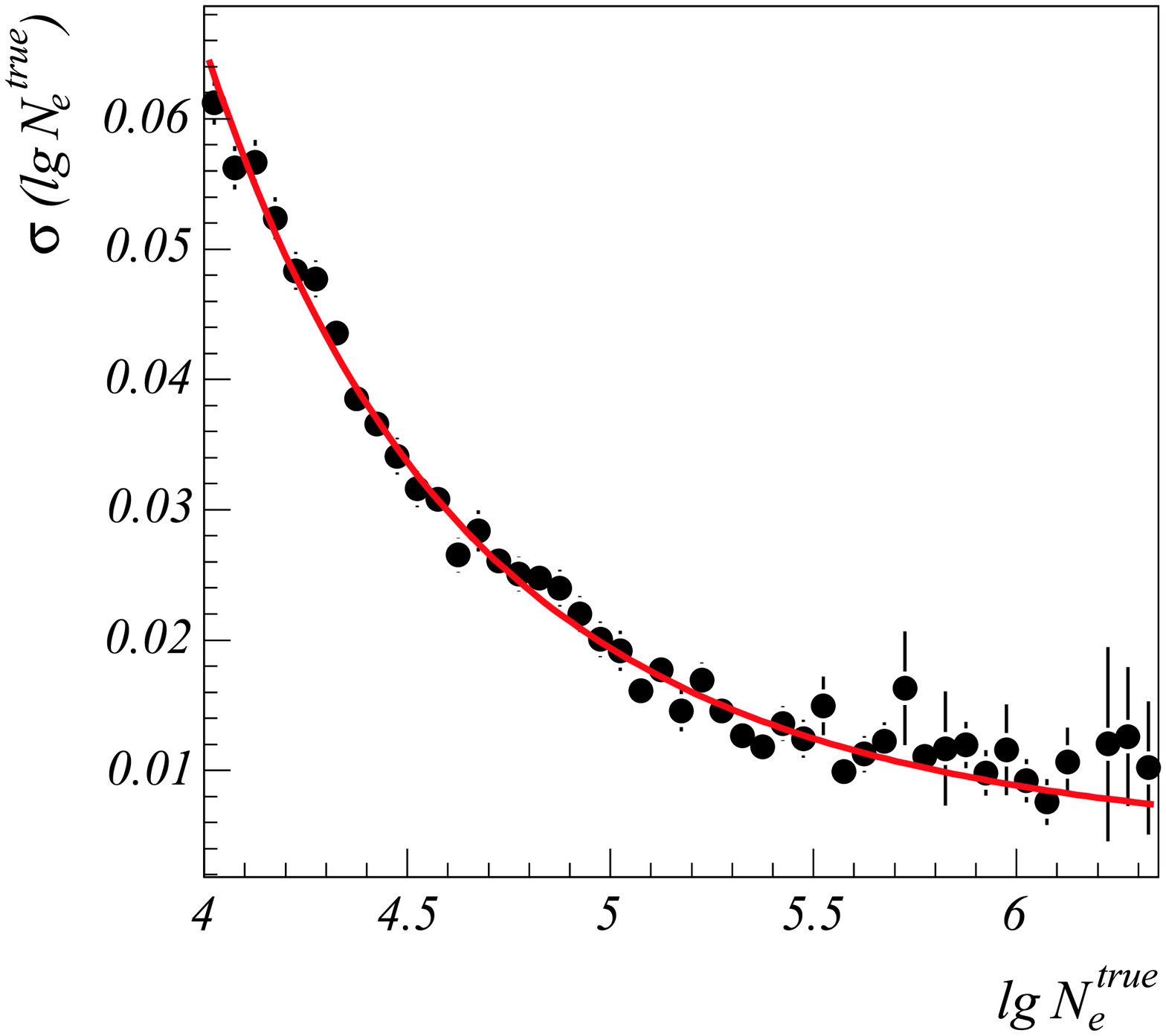,width=\linewidth}
\end{minipage}
\caption{Left: Distribution of $\lg N_{e}-\lg N_{e}^{true}-C_{e}$ for proton
induced showers with $4.2<\lg N_{e}^{true}\leq4.3$ fitted with a Gaussian
distribution.
Right: Dependence of the width $\sigma$ on the true electron number
$\lg N_{e}^{true}$.}
\label{elerek_resol}
\end{figure}

The dependence of the corrected difference
$\lg N_{e}-\lg N_{e}^{true} - C_{e}$
on true electron number is displayed in the right part of Fig.
\ref{elerek_syst}.
The increase towards low values of $\lg N_{e}^{true}$ is due to the combined
threshold of measurement and reconstruction, the decrease of
$\lg N_{e} - \lg N_{e}^{true} - C_{e}$ towards larger shower sizes reflects
saturation effects in the array detectors which influence the quality of the
reconstruction. These deviations from the zero line are parameterized and
accounted for in the analysis.

The distribution of $\lg N_{e} - \lg N_{e}^{true} - C_{e}$ for fixed
true electron number $\lg N_{e}^{true}$ can be described with good quality
by a Gaussian. An example for this is displayed in the left part
of Fig.~\ref{elerek_resol}. The form and the parameters of this distribution
do not depend on primary particle type.
The width of the distribution depends on
$\lg N_{e}^{true}$ but can be easily parameterized which is displayed in
the right part of Fig.~\ref{elerek_resol}.
The adopted description of the reconstruction systematics and resolution
of $\lg N_{e}$ can be integrated into the calculation of
the response matrix elements.
The influence of resolution effects on the results are small since for the
showers used ($\lg N_{e}>4.8$) the
resolution is smaller than the bin width in $\lg N_{e}$.

\begin{figure}[t]
\begin{minipage}[t]{.50\linewidth}
\centering\epsfig{file=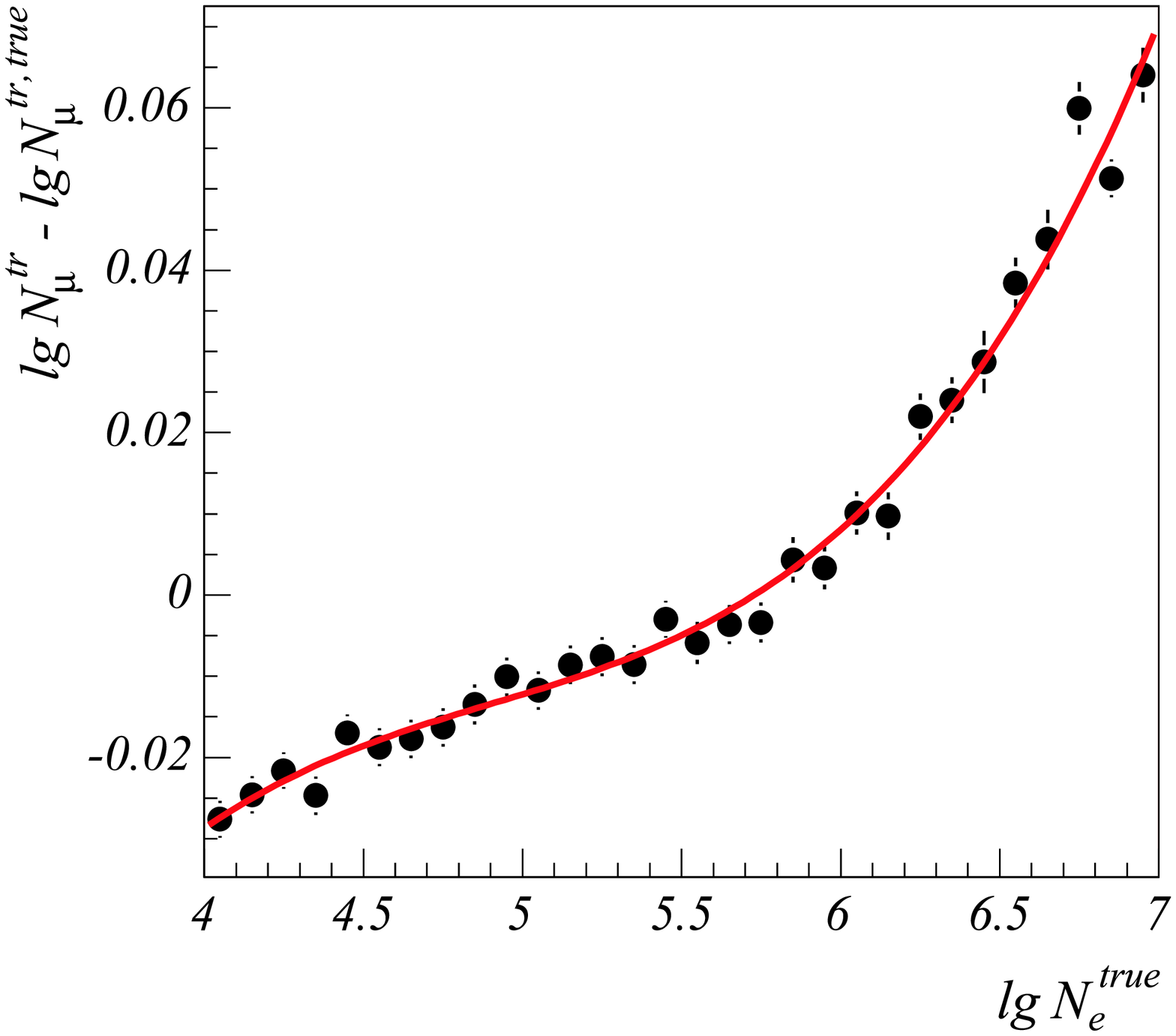,width=\linewidth}
\end{minipage}
\begin{minipage}[t]{.50\linewidth}
\centering\epsfig{file=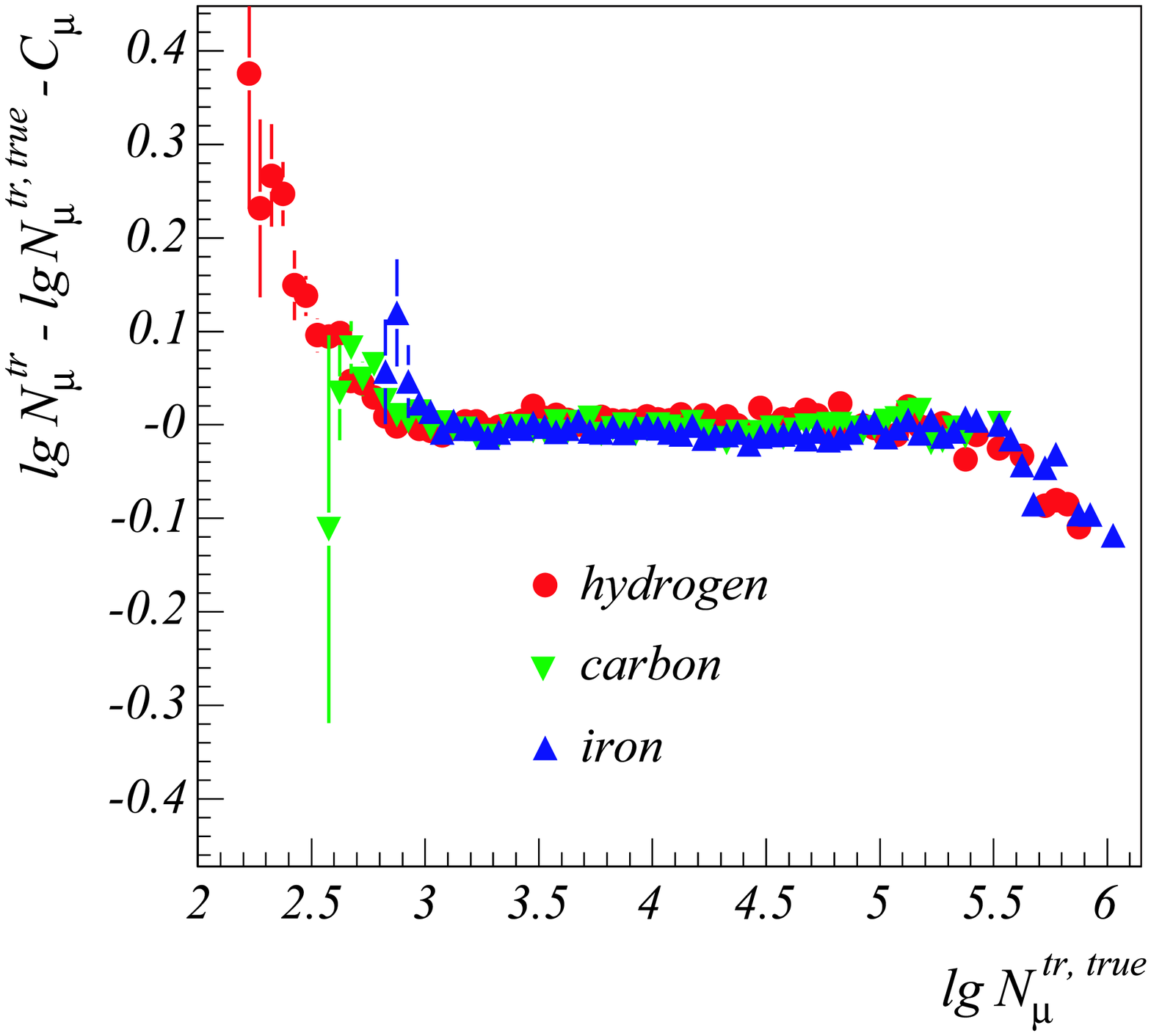,width=\linewidth}
\end{minipage}
\caption{Left: Relation between $\lg N_{\mu}^{tr}-\lg N_{\mu}^{tr,true}$
and $\lg N_{e}^{true}$ used for the correction of $\lg N_{\mu}^{tr}$.
Right: Remaining systematic differences between reconstructed and true
truncated muon number after correction with the relation in the left part of
the figure.}
\label{myorek_syst}
\end{figure}

In the case of the truncated muon number a correlation between the difference
$\lg N_{\mu}^{tr} - \lg N_{\mu}^{tr,true}$ and the true electron
number $\lg N_{e}^{true}$ was found. This correlation, displayed in the left
part of Fig.~\ref{myorek_syst}, proved also to be nearly independent of
primary particle type.
Using this correlation for the parameterization of a correction term $C_{\mu}$
any dependence of
$\lg N_{\mu}^{tr} - \lg N_{\mu}^{tr,true}$ on $\lg N_{\mu}^{tr}$ nearly
vanishes. The mean values of $\lg N_{\mu}^{tr} - \lg N_{\mu}^{tr,true}
- C_{\mu}$ versus muon number $\lg N_{\mu}^{tr,true}$ are displayed in the
right
part of Fig.~\ref{myorek_syst}. Deviations from the zero line for small and
large values of the muon number have similar reasons as in the case of
$\lg N_{e}$ and are accounted for.

\begin{figure}[b]
\begin{minipage}[t]{.50\linewidth}
\centering\epsfig{file=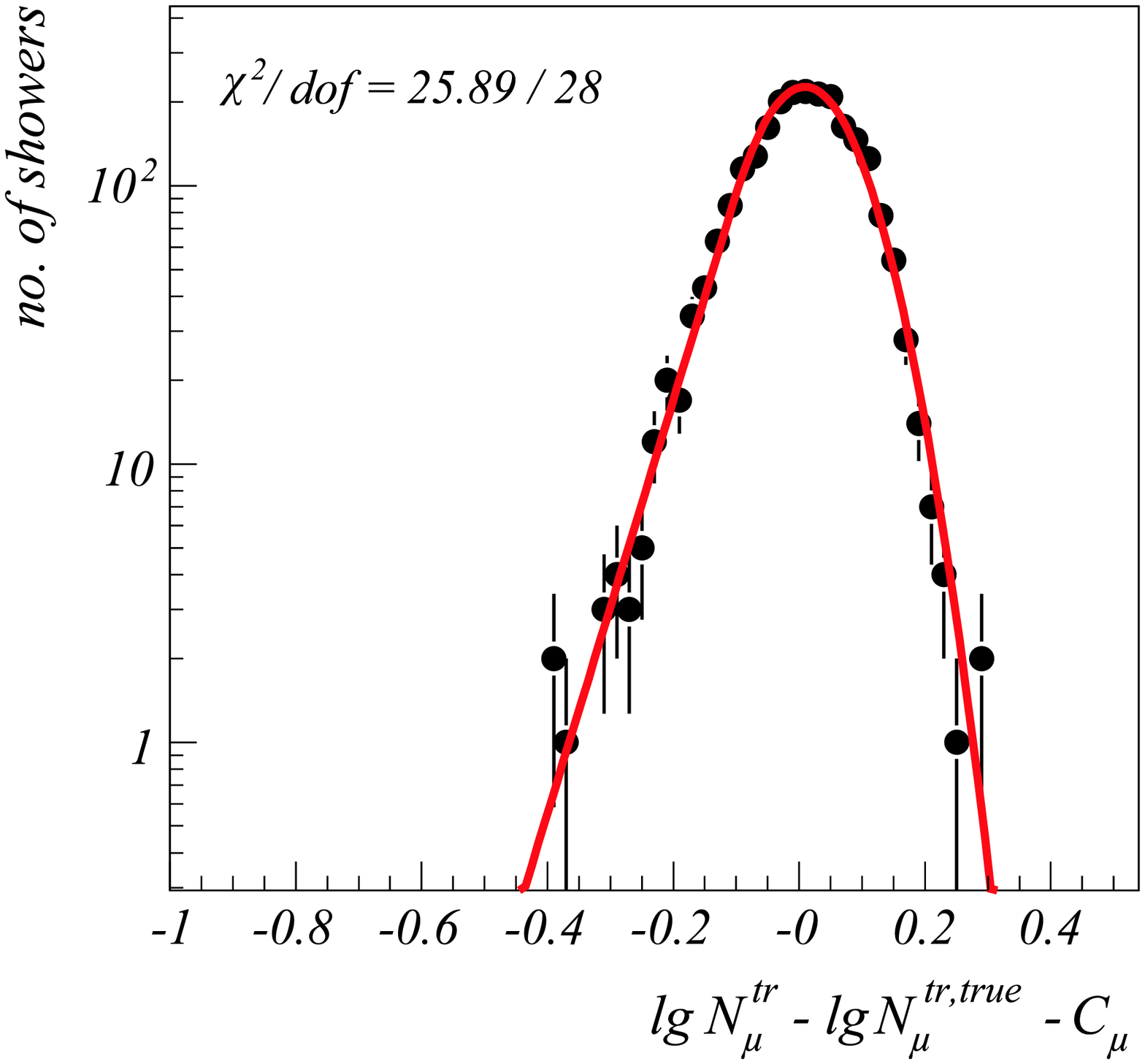,width=\linewidth}
\end{minipage}
\begin{minipage}[t]{.50\linewidth}
\centering\epsfig{file=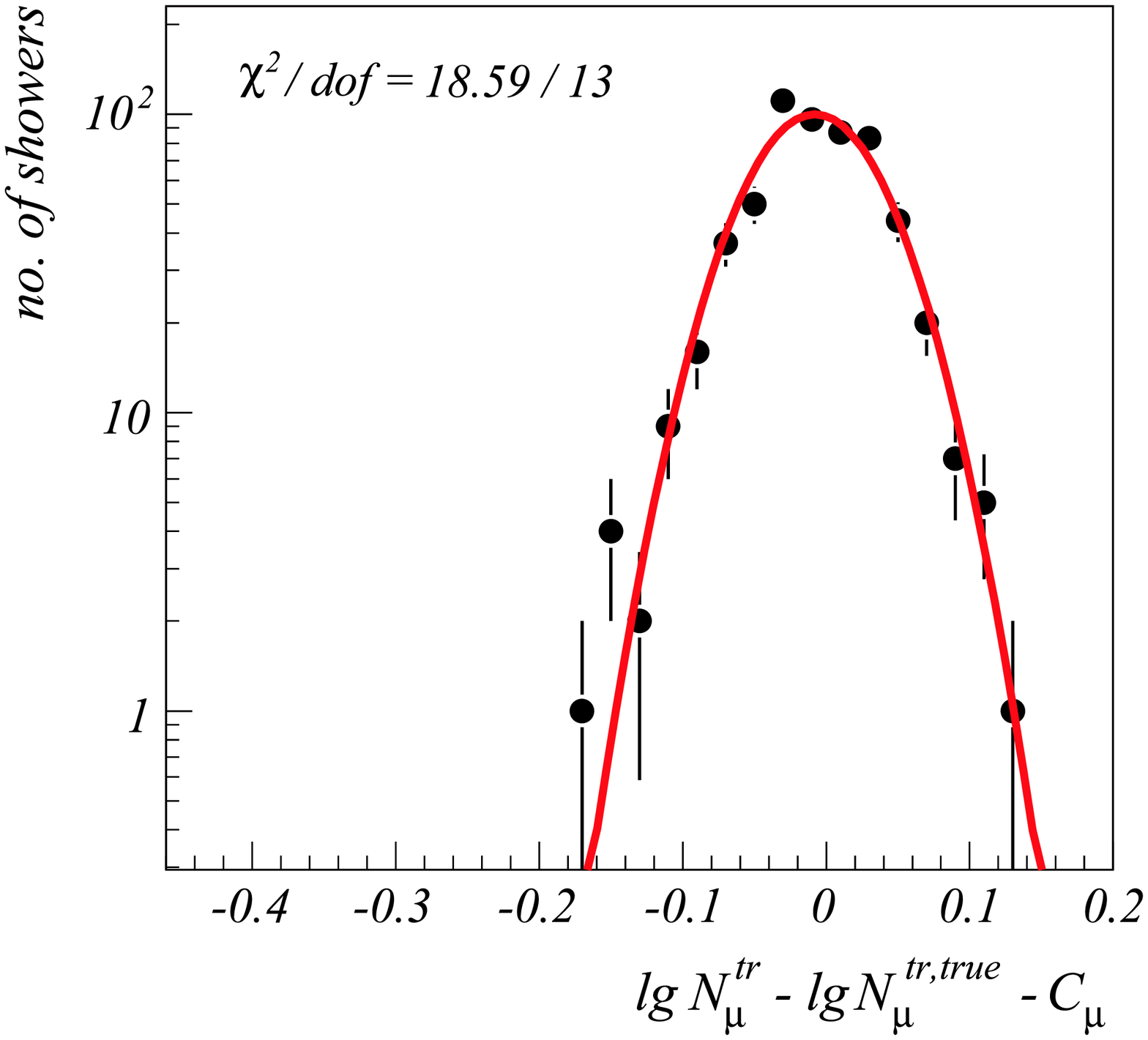,width=\linewidth}
\end{minipage}
\caption{Distribution and fit of $\lg N_{\mu}^{tr}-
\lg N_{\mu}^{tr,true}-C_{\mu}$ for showers with
$3.6<\lg N_{\mu}^{tr,true}\leq3.7$ (left) and
$4.2<\lg N_{\mu}^{tr,true}\leq4.3$ (right).}
\label{myorek_resol}
\end{figure}
\begin{figure}[t]
\begin{minipage}[t]{.50\linewidth}
\centering\epsfig{file=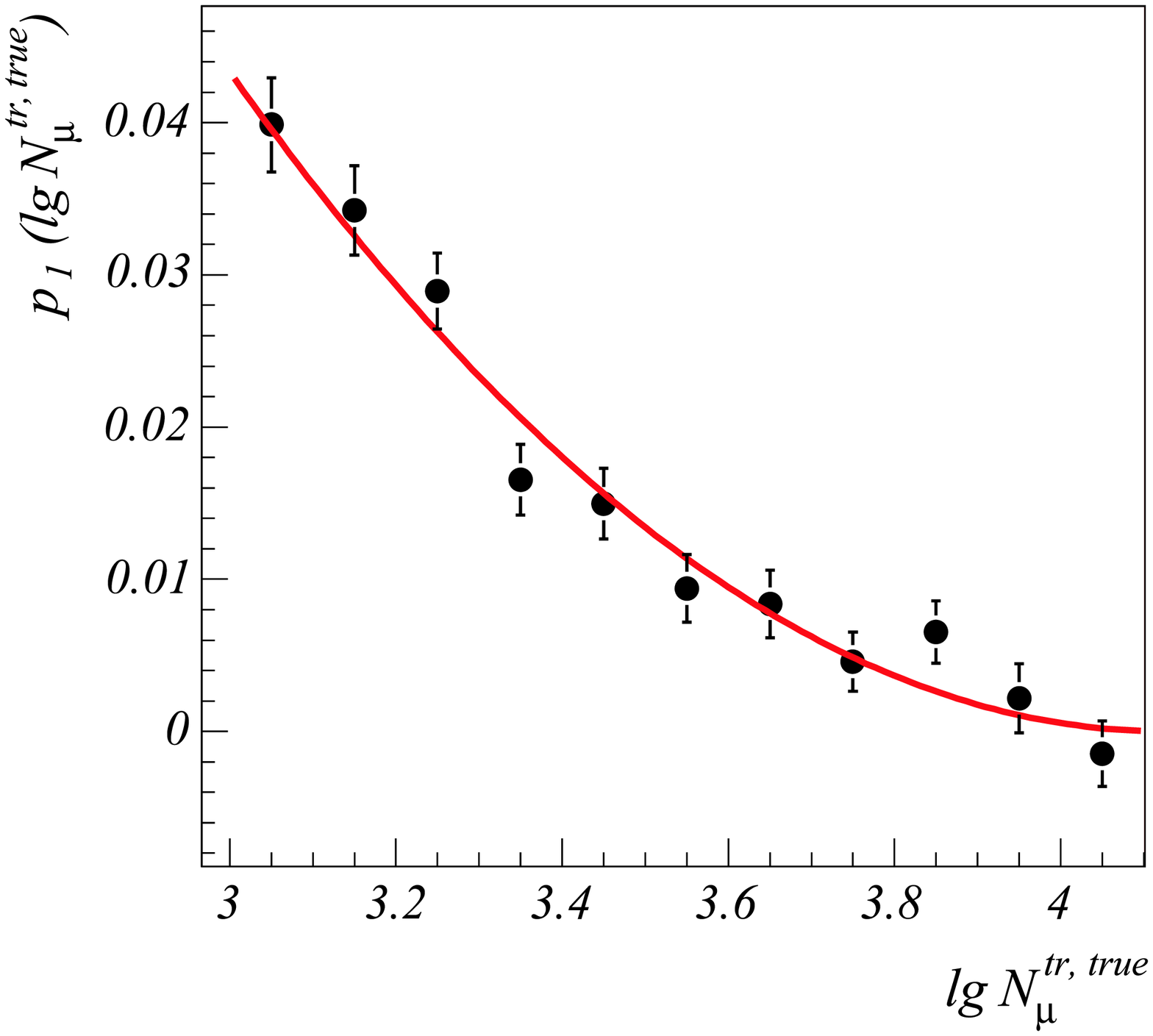,width=\linewidth}
\end{minipage}
\begin{minipage}[t]{.50\linewidth}
\centering\epsfig{file=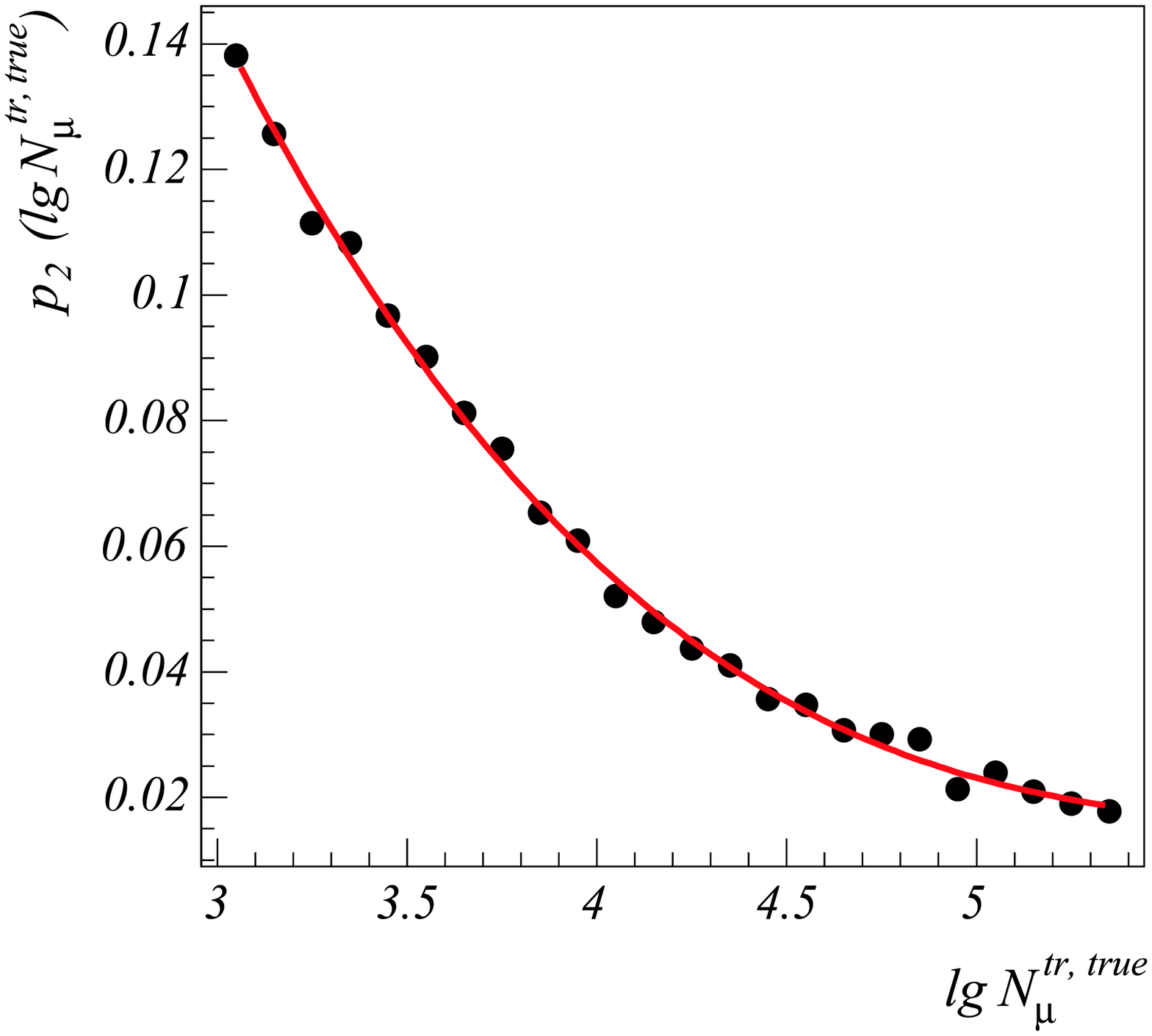,width=\linewidth}
\end{minipage}
\caption{Dependence of parameters $p_{1}$ (left) and $p_{2}$ (right) of
Eq.~(\ref{muon_streu}) on the true muon number $\lg N_{\mu}^{tr,true}$.}
\label{myorek_paras}
\end{figure}
In contrast to the case of the electron number the distribution of
$\lg N_{\mu}^{tr} - \lg N_{\mu}^{tr,true} - C_{\mu}$ is asymmetric for smaller
values of $\lg N_{\mu}^{tr,true}$ but becomes more and more symmetric with
increasing $\lg N_{\mu}^{tr,true}$. For values $\lg N_{\mu}^{tr,true}>4$
the distribution can be described again by a Gaussian. In Fig.
\ref{myorek_resol} these
distributions are displayed for two narrow intervals of
$\lg N_{\mu}^{tr,true}$. In order to describe the asymmetric distribution the
following functional form was used
\begin{equation}
r(x) = \bigg\{ \begin{array}{ll}
c_{1} \cdot e^{- \frac{(x-p_{1})^{2}}{2\cdot p_{2}^{2}}} & \quad
x > p_{1} - \frac{p_{2}^{2}}{p_{3}}\\
c_{1}c_{2} \cdot e^{\frac{x}{p_{3}}} & \quad x \leq p_{1} -
\frac{p_{2}^{2}}{p_{3}} \end{array}
\label{muon_streu}
\end{equation}
with the factor $c_{2}$ as a normalization constant. The value of $p_{1}$ tends
to zero with increasing $\lg N_{\mu}^{tr,true}$. 
The dependence of $p_{1}$ and $p_{2}$,
which can be considered as a measure for the resolution of the muon number
determination, on the true muon number are displayed in Fig.
\ref{myorek_paras}.
\section{Solving the matrix equation}\label{strategy}
\subsection{Application of unfolding methods}
From a purely mathematical point of view, solving the matrix equation
Eq.~(\ref{simplemateq}), only requires a simple inversion of the
matrix ${\bf R^{-1}}$.
However, a closer inspection of this matrix shows,
that it is close
to singularity and Eq.~(\ref{simplemateq}) states an ill-conditioned problem.
Therefore, a direct inversion would give meaningless results.

The reason for the poor condition of the response matrix is closely
related to the properties of shower fluctuations.
Since for different primary particles the corresponding distributions
overlap to a large extent, the
discrimination between these particles gets more and more difficult with
increasing
number of particle types. In the extreme case of very similar particles,
like for example nitrogen and oxygen, the corresponding matrix elements
would coincide inside the computational accuracy for a
reasonable binning of the data. In this case ${\bf R}$ is singular,
and an inversion impossible. Therefore, the number of considered particle
types has to be restricted.

Another reason lies in the steeply falling primary spectrum.
Due to their broad shower fluctuations low 
energy primary particles may be registered at high electron and 
muon numbers.
Although the probabilities for this are very small this may be
compensated by their high flux. 
This is reflected by a few very small matrix elements
(in the order of $10^{-5}$ and smaller). As a result, nearly
identical rows and columns consisting of very small values are present in
${\bf R}$ even when only one primary particle type is considered.
Again, this leeds to a nearly singular matrix.
In addition, also many small off-diagonal elements are introduced in the
response matrix which are sensitive to rounding errors.

Altogether, the response matrix
${\bf R}$ exhibits nearly identical rows and columns and many small
off-diagonal elements. In such case, inversion of a matrix is in general an
ineffective strategy, and one
has to rely
on methods which approximate the solution avoiding the problems
inherent to matrix inversion.
One class of methods especially suited for the determination of approximate
solutions of ill-conditioned matrix and integral equations are
so-called {\it unfolding} or {\it deconvolution} techniques. For this there
exist many different algorithms, each one with its own systematic properties.
To get a measure for the size of the
systematic errors caused by the unfolding three different methods are used
in the present analysis.
These are the Gold algorithm \cite{gold}, unfolding based
on the Bayesian theorem \cite{bayes} and an entropy based unfolding method
\cite{schmelling}. Characteristic of these procedures is the
generation of only non-negative
solutions. Properties of these methods and details about their
application are briefly presented in Appendix \ref{unfold_app}.
\subsection{Considered primary elements}
For unfolding techniques to be applicable the matrix equation 
(\ref{simplemateq}) has to exhibit a minimum degree of stability 
for the algorithms to provide meaningful solutions.
This stability is characterized by the condition number of the response matrix,
which is strongly influenced by the number of
primary particle types included in the analysis.
One is restricted to a maximum
number of elements since otherwise this would lead to a singular matrix.
The relevant quantity here is the  {\it condition number} 
defined by the ratio of
the biggest to the smallest singular value of a matrix. The larger its
value, the poorer is the condition of the matrix. Acceptable values
are in the range of $10^{6}$ to $10^{7}$, depending on the the specific
problem.

To find maximum number of particle types, the number 
was varied, and in each case an unfolding procedure performed.
For this investigations results of QGSJet 01 based simulations were used.
The quality of the results was judged by means of a $\chi^{2}$-comparison
with the measured data (see section \ref{qgsdiscuss} or \ref{sibdiscuss} for
details). To determine the condition number of the corresponding matrices a
singular value decomposition was performed.
For the use of only two primaries (H and Fe) a $\chi^{2}$
per degree of freedom of 245 was achieved, for three particles (H, C, Fe) a
value of 35, for four particles a value of 3.3 in the case of H, He, C, and Fe,
and 2.5 for the use of H, He, C and Si, respectively. For five elements
(H, He, C, Si, Fe)
a value of 2.38 for $\chi^{2}$ per degree of freedom was found.
At the same time the condition
number increases from $2.3\cdot10^{5}$ (H, Fe) to $1.7\cdot10^{6}$ (H, C, Fe)
and $4\cdot10^{6}$ for four primaries up to $8.5\cdot10^{6}$ in the case
of H, He, C, Si and Fe. In addition, a significant increase in
the statistical uncertainty of the solution with the transition from four
to five primaries was observed.

Due to the already large condition number in the case of five elements and
only small improvement in the description of the data by the solution,
finally five primary particle types are adopted for the analysis. These
are hydrogen (protons), helium, carbon, silicon, and iron.
The spectra of proton and helium will describe the energy spectra of
single elements, whereas the
three other types represent elemental groups only,
carbon essentially the CNO-group, silicon the intermediate, and iron the
heavy elements. Furthermore, it is not possible to specify from which
elements of these groups the resulting energy spectra stem.
\section{Monte Carlo tests}\label{decexamp}
Before applying any of the unfolding algorithms to measured data
it has to be tested if the method is suited for the actual problem. In
order to get an estimate of the capabilities and sensitivity of an
algorithm it is tested in an ``ideal environment''.
According to an assumed energy spectrum for each primary particle type
energy values are randomly chosen. Electron and muon numbers for
each energy value are generated by Monte Carlo techniques, 
using the parametrized shower
fluctuations and reconstruction properties, 
and a two-dimensional shower size spectrum,
in range and binning identical with the measured one, is filled. The
generated data correspond to approximately one third of the KASCADE data
used. This artifical data histogram is input for the unfolding algorithms.

\begin{figure}[t]
\begin{minipage}[b]{.50\linewidth}
\centering\epsfig{file=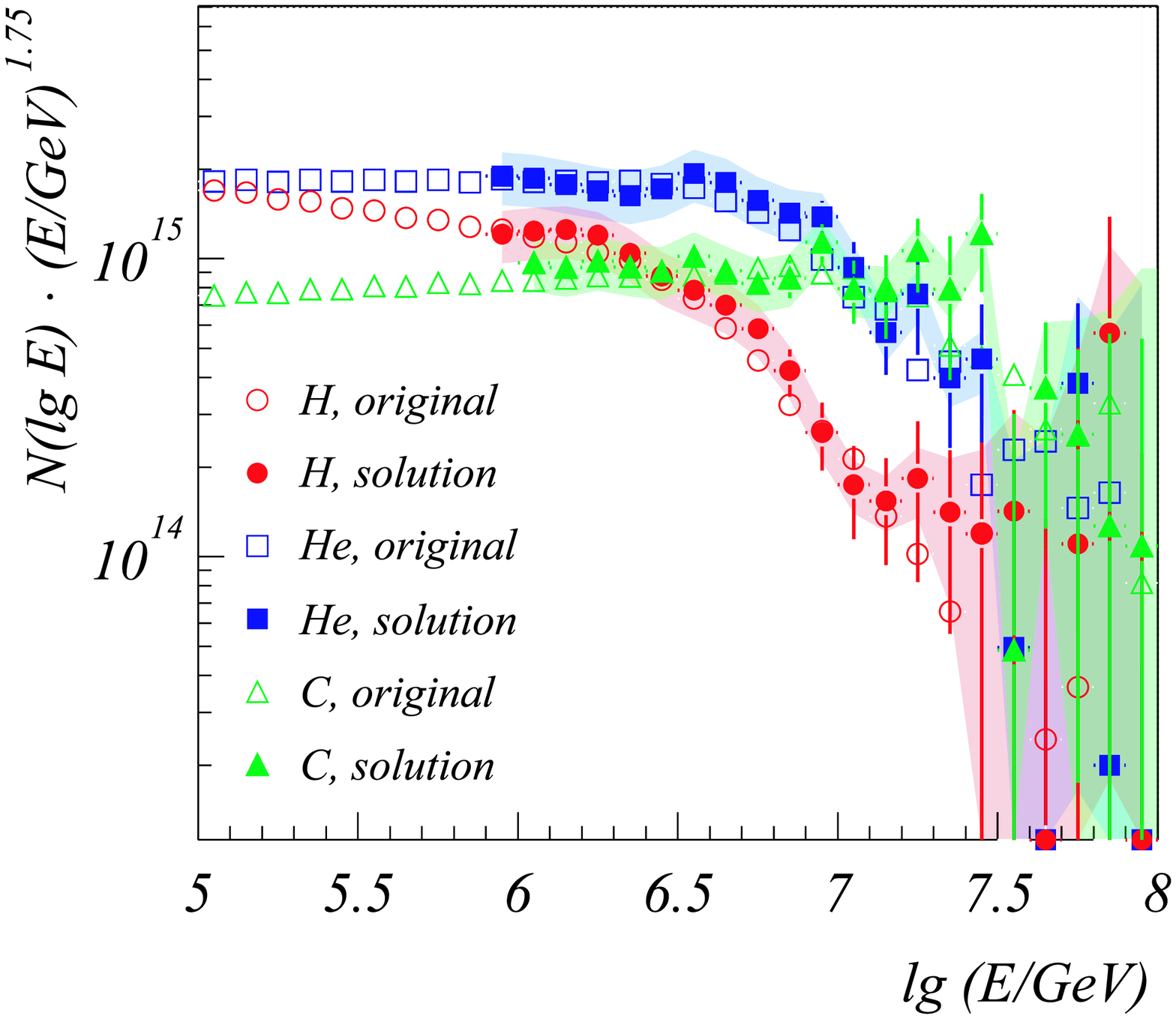,width=\linewidth}
\end{minipage}
\begin{minipage}[b]{.50\linewidth}
\centering\epsfig{file=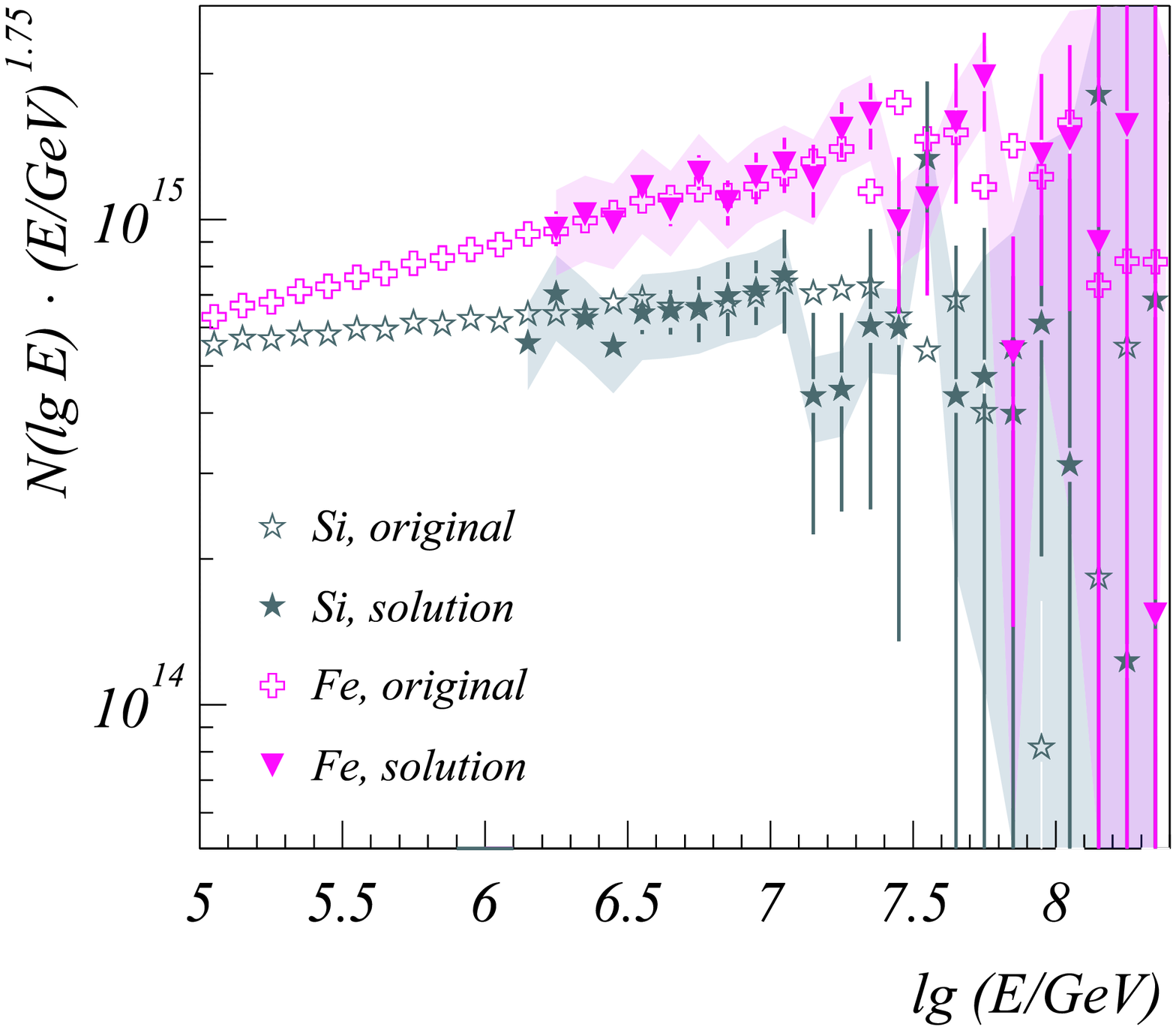,width=\linewidth}
\end{minipage}
\caption{Unfolding results (filled symbols) for the energy spectra of H, He
and C (left panel) and Si and Fe (right panel) together with the original
"true" spectra (open symbols). The shaded bands are an estimate of the
systematic uncertainties due to the applied unfolding method, in this case
the Gold algorithm.}
\label{MCunfoldtest}
\end{figure}

The test procedure was carried out for different
assumptions of the individual energy spectra. Considered cases include knee
features in each spectrum at different energies, knee features at the
same primary energy, knee features only in some of the elemental spectra, and
only simple power laws (no knees at all). In all cases similar good
results were achieved. In the following the test procedure and its results
are presented for the example of a rigidity dependent knee.
The assumed energy spectra follow a power law
exhibiting a knee with the individual knee positions chosen to be
proportional to the particle charges.
All three unfolding methods yielded
good and comparable results. As an example,
the results of the Gold algorithm is 
presented here which is preferred because of its speed
and robustness.
A comparison between the results of the different unfolding algorithms
is presented in section~\ref{theqgsresults} and in Fig.~\ref{QGSunfoldcomp}
for the unfolding of the KASCADE data using QGSJet simulations.
The "true" spectra of proton, helium, carbon,
silicon, and iron are depicted by the open symbols in Fig.~\ref{MCunfoldtest}.
Flux values and spectral indices below the knee are based on the
compilation of \cite{wiebel}.

To obtain reliable results, some criterion to
stop the iteration is required. For the determination of the
adequate number of iteration steps the weighted mean squared error
(WMSE) and the relative variance of the bias (RBS) are used. These quantities
are defined by
\begin{equation}
{\mathrm{WMSE}} = \frac{1}{m}\sum_{i=1}^{m}\frac{\sigma_{X,i}^{2}+b_{i}^{2}}
{\tilde{X}_{i}}\quad{\mathrm{and}}\quad {\mathrm{RBS}} =
\frac{1}{m}\sum_{i=1}^{m} \frac{b_{i}^{2}}{\sigma_{b,i}^{2}},
\end{equation}
where $m$ is the dimension of the solution, $\tilde{X}_{i}$ the
value of the $i$th element of the solution vector $\vec{X}$ and
$\sigma_{X,i}^{2}$ its variance; $b_{i}$ is the systematic bias from the true
value and $\sigma_{b,i}$ the statistical uncertainty of this bias.
Details about their use in unfolding analyses can be found in \cite{cowan}.
For the determination of the WMSE and RBS a bootstrap method is used. The
solution of the current iteration step serves as model for the generation
of a set of Monte Carlo data which are deconvoluted. The obtained
solution set is compared to the input spectra, i.e. the original 
solution. This method proved to work well for the 
estimation of the WMSE and
RBS but provides a good estimate of the absolute value of the average
systematic uncertainties only, and not for their sign.

The result of the unfolding is shown in Fig.~\ref{MCunfoldtest} for 
the Gold method. The open symbols
correspond to the original ``true'' spectra, filled symbols represent the
solution of the unfolding procedure. The left part of the figure displays the
spectra of protons, helium, and carbon, the spectra of silicon and iron are
shown in the right part. Error bars represent statistical errors whereas
systematic uncertainties are shown as shaded bands.
Statistical errors are due to the limited number of simulated data
and are estimated by repeating the unfolding with different sets. Systematic
uncertainties are estimated by comparison of the mean value of the
set of results with the values of the original ``true'' spectra.
These systematic uncertainties are mainly due to the uncertainty
in terminating the iteration and the value of the regularization parameter,
respectively. For all three unfolding methods this uncertainty is of
order 15\% for low energies, i.e. high fluxes.
The strong increase of the systematic uncertainties at higher energies is
due to the low fluxes and hence small number of showers.
Since the considered algorithms are designed to generate only
non-negative solutions they tend to introduce an additional bias in the case
of small number of events. This bias gets significantly large
for energies with less than $\approx30$ events per bin.

As can be seen in Fig.~\ref{MCunfoldtest}, the spectral features of the
original spectra,
like knee position and spectral index, are well reproduced
within the
statistical and systematic uncertainties. The artifical
``wobbling'' at low energies can be identified as a systematic
relic of the unfolding procedure. Assuming a smooth behaviour of the
original spectrum this yields an independent estimate for the size of the
systematic uncertainties, again of order 15\%. For the determination of the
spectral shapes and
the spectral indices these systematic effects have to be considered.
Altogether,
it can be concluded that the proposed analysis technique is applicable to the
problem of unfolding the two-dimensional air shower size spectrum with five
primary mass groups.
\clearpage
\section{Results}\label{results_sec}
\subsection{Results based on QGSJet 01}\label{theqgsresults}
\begin{figure}[b]
\centering\epsfig{file=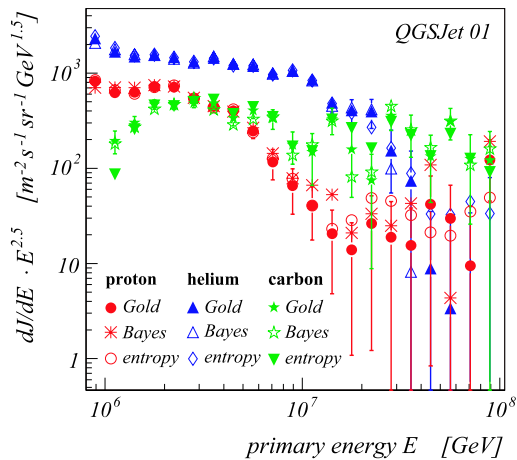,width=0.5\linewidth}
\caption{Results using QGSJet hypothesis for the elements H, He, C and for
three different unfolding algorithms. For reason of clarity statistcal
error bars are displayed for the results of the Gold algorithm only.}
\label{QGSunfoldcomp}
\end{figure}

All three unfolding procedures mentioned
above were used in order to cross-check the solution. The result
for the energy spectra of the light element groups (H, He, C) are shown in
Fig.~\ref{QGSunfoldcomp}
for the three methods. The different unfolding results agree very well with
each other. The same holds also for the heavy groups and also for the
results based on SIBYLL simulations, presented in
section~\ref{thesibyllresults}.
Therefore, in the
following only the results using the Gold algorithm will be discussed 
which showed the highest
speed and robustness among the applied methods.
\begin{figure}[t]
\centering\epsfig{file=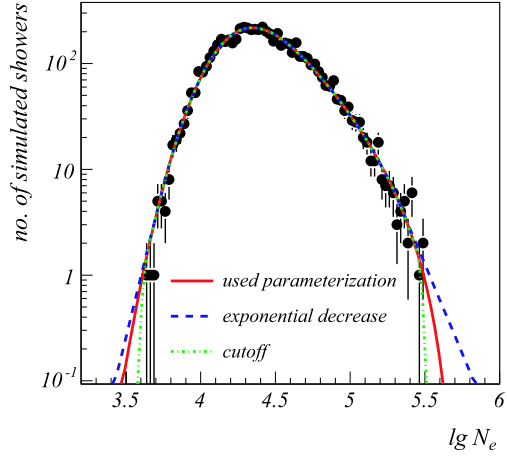,width=0.5\linewidth}
\caption{Different
extrapolations of the $\lg N_{e}$-distribution for 0.5~PeV proton
induced showers (QGSJet 01).}
\label{diffparameteriz}
\end{figure}

In addition to the statistical uncertainties due to
the limited number of measured showers and the systematic uncertainties due
to the unfolding
algorithm, two additional sources of uncertainties have to be considered.

First, the number of simulated showers is limited, giving rise to further
statistical uncertainties of the fit parameters. To estimate this
influence, each parameter of Eq.~(\ref{correlprobab}) is varied randomly
within its error distribution. For each new set of parameters 
the energy dependence
is interpolated and new response matrices are calculated. The 
unfolding is repeated with each set of
response matrices  and the spread of the
individual fluxes determined. This additional statistical error is already
included in the error bars in Fig.~\ref{QGSunfoldcomp}.
\begin{figure}[b]
\begin{minipage}[b]{.50\linewidth}
\centering\epsfig{file=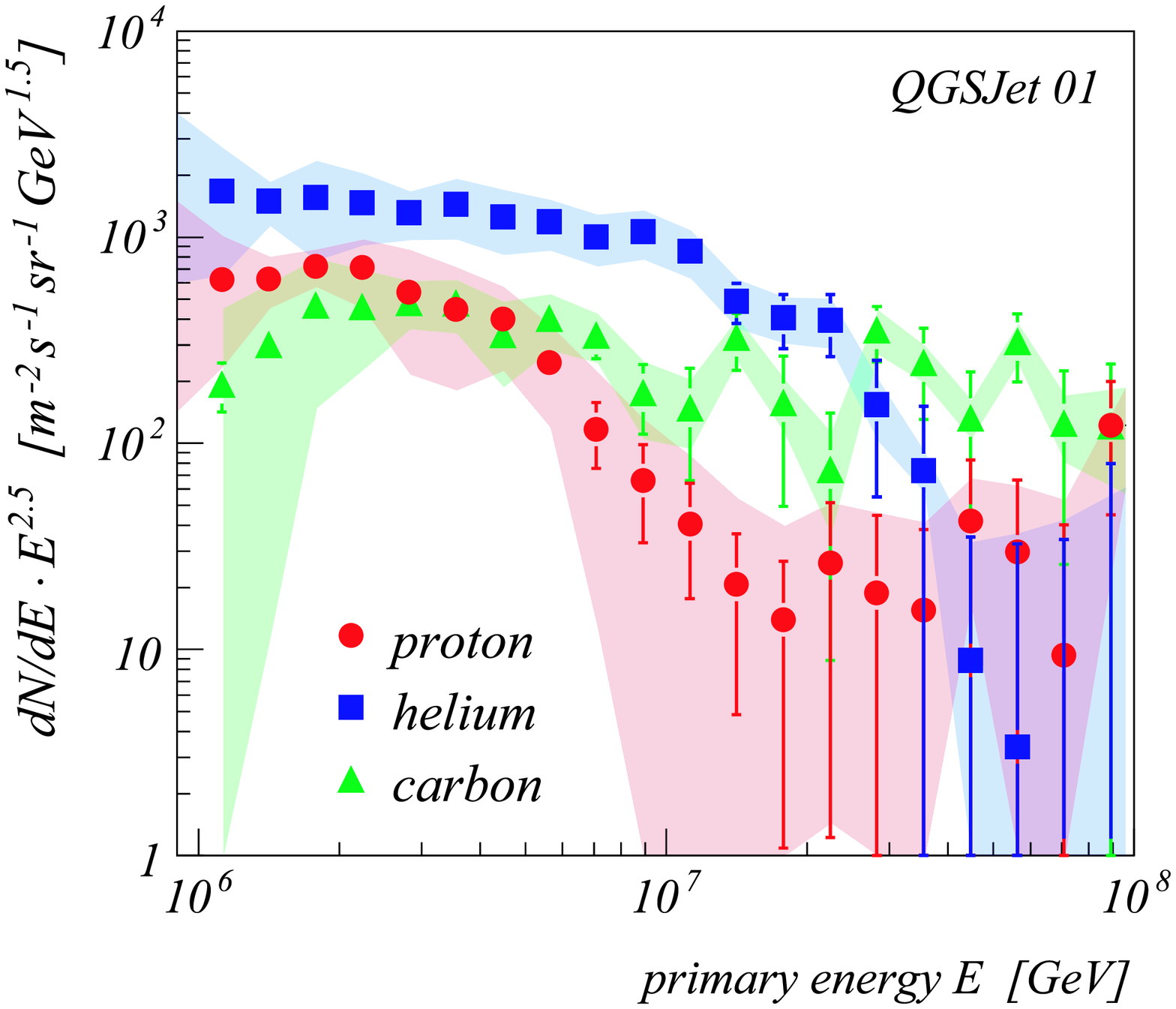,width=\linewidth}
\end{minipage}
\begin{minipage}[b]{.50\linewidth}
\centering\epsfig{file=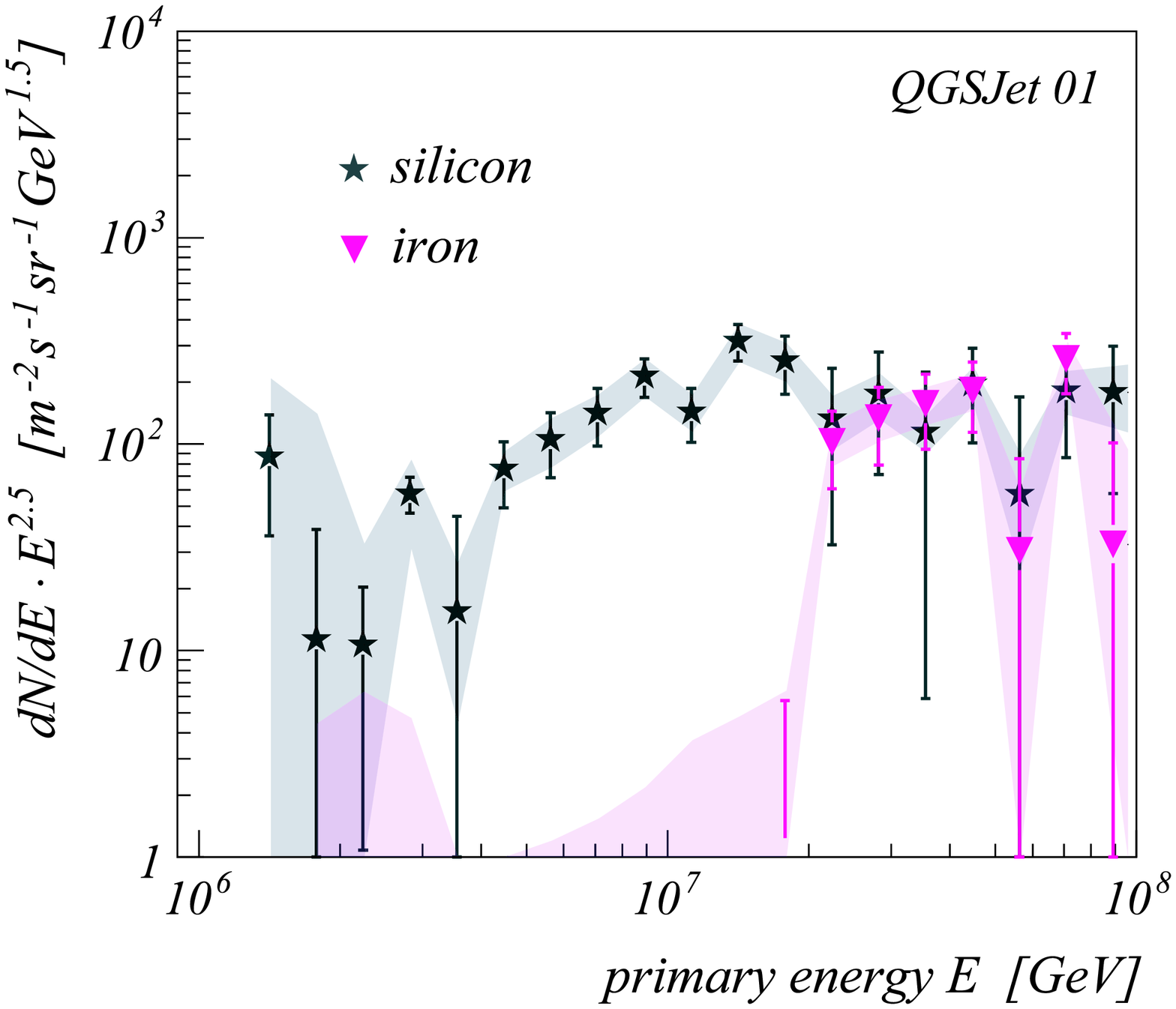,width=\linewidth}
\end{minipage}
\caption{Unfolded energy spectra for H, He, C (left panel) and Si, Fe (right
panel) based on QGSJet simulations. The shaded bands are an estimate of the
systematic uncertainties due to the used parametrizations and the applied
unfolding method (Gold algorithm).}
\label{QGS_unfoldbands}
\end{figure}

Second, the form of the tails of the shower size distributions is not known.
Fig.~\ref{diffparameteriz} shows an example of the
$\lg N_{e}$--distribution for showers induced by 0.5~PeV protons.
Besides the parameterization used, two different
extrapolations are displayed, the first one with 
sharp cutoffs at the edges of
the distribution, the second one with an exponential decrease up
to higher and lower values of $\lg N_{e}$. Within the
statistics of the simulations 
each of these functions describes the distribution equally well.
The influence of these tails on the shower
size spectra and the unfolding result may be quite important because of the
steeply falling primary energy spectra.
The displayed parameterizations in Fig.~\ref{diffparameteriz} can be
regarded as extreme assumptions and it has been investigated that the
corresponding unfolding results form an upper and lower bound for the spectra.
This range can be considered as an
estimate for the systematic uncertainty due to the unknown shape of the
distribution tails.
It should be mentioned that the size of this systematic uncertainty 
should, according to simulations, be considerably reduced for 
observations close to shower maximum (e.g. around 5000~m a.s.l.).

In Fig.~\ref{QGS_unfoldbands} the unfolding
result is displayed together with the estimate of the total systematic
uncertainty, shown as shaded bands. For low energies, the dominant
contribution to the systematic uncertainty
is due to the tails of the distributions.

Below the knee helium is the most abundant element, followed by
protons and carbon. The energy spectra of both proton and helium show a
knee-like feature whereas for
carbon no knee structure is visible. The spectra of the heavier elements
look rather unexpected, especially in the case of iron. For energies below
10~PeV practically no iron is present, above 20~PeV it dominates
the cosmic ray spectrum together with silicon.
\subsection{Results based on SIBYLL 2.1}\label{thesibyllresults}
\begin{figure}[b]
\begin{minipage}[b]{.50\linewidth}
\centering\epsfig{file=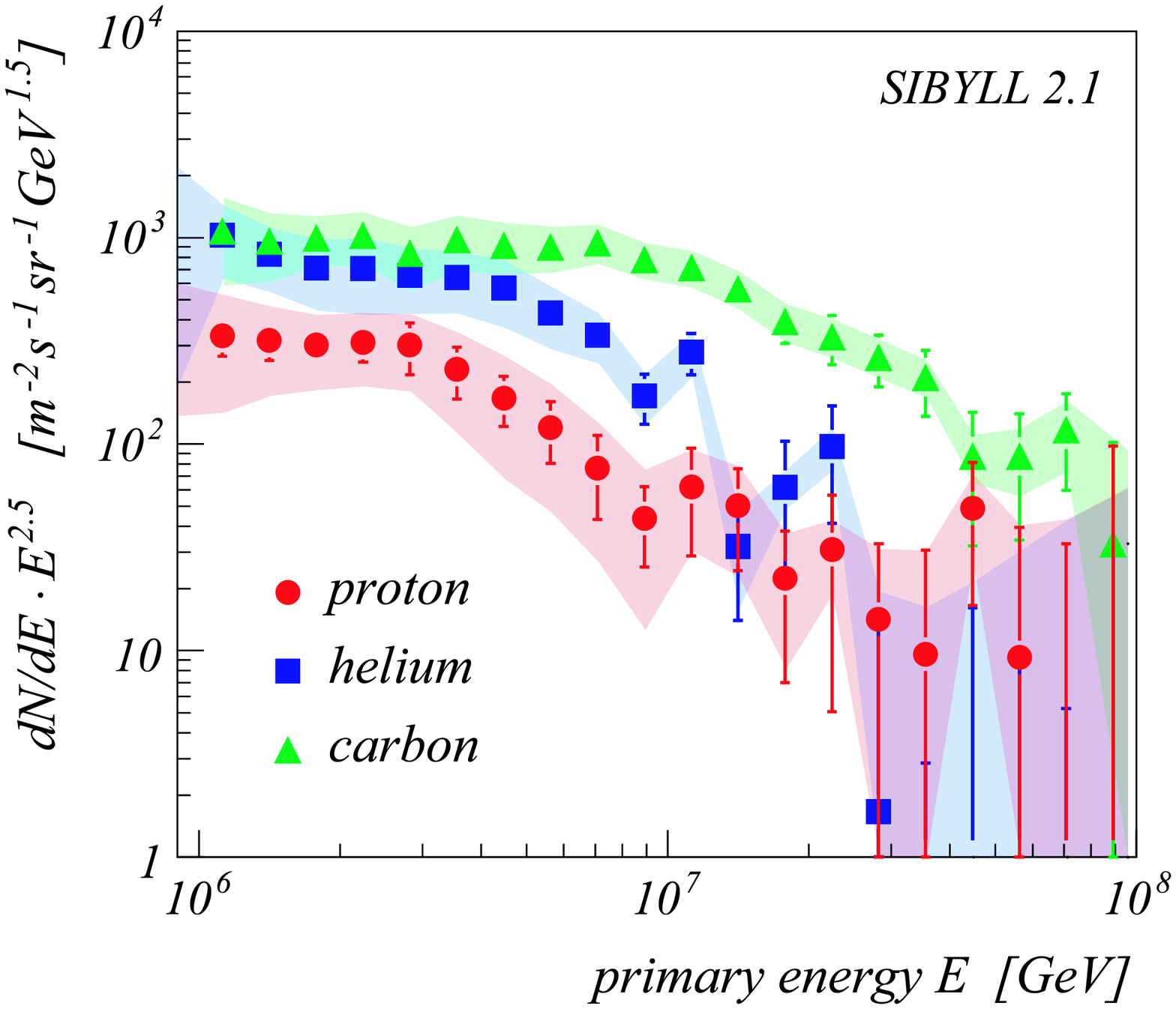,width=\linewidth}
\end{minipage}
\begin{minipage}[b]{.50\linewidth}
\centering\epsfig{file=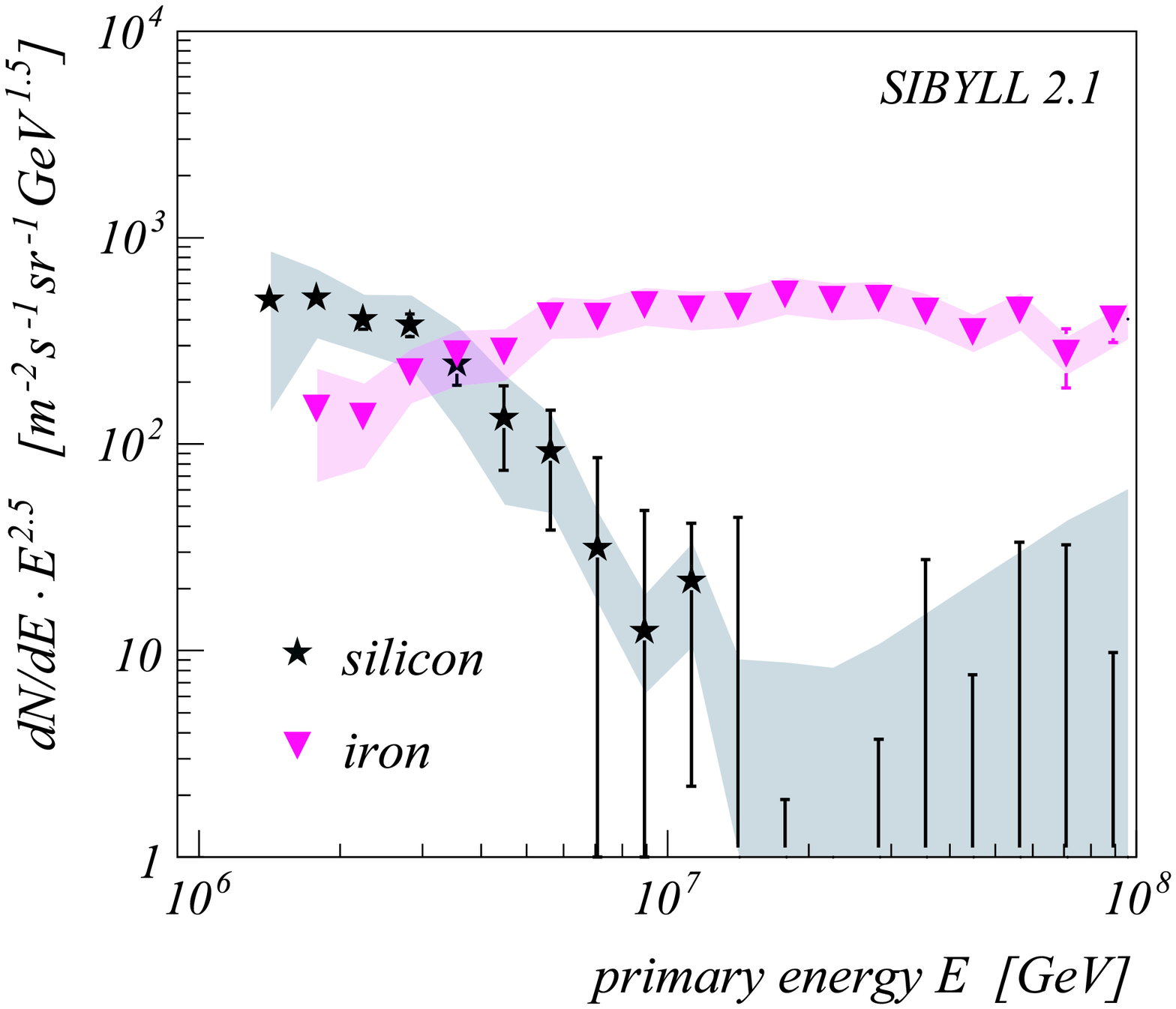,width=\linewidth}
\end{minipage}
\caption{Unfolded energy spectra for H, He, C (left panel) and Si, Fe (right
panel) based on SIBYLL simulations. The shaded bands are estimates of the
systematic uncertainties due to the used parameterizations and the applied
unfolding method (Gold algorithm).}
\label{SIB_unfoldbands}
\end{figure}

The outcome of the unfolding
using CORSIKA/SIBYLL/GHEISHA for calculation of the response matrices is
presented in Fig.~\ref{SIB_unfoldbands} for the Gold algorithm 
and five particle types. 
As in the case of the QGSJet analysis the different unfolding methods give
essentially equal results.
The estimated total systematic
uncertainties at lower energies are slightly smaller than for the
QGSJet based results
due to a better description
of the measured data in the corresponding data range, which will be
discussed in section~\ref{sibdiscuss}.
Each of the spectra of the light
groups (proton, helium and CNO) shows a knee-like feature. The position
of the individual knees is shifted to higher energies with increasing
atomic number. In contrast to the QGSJet results, carbon is the 
most abundant element
at energies around 1-2~PeV but helium is again more abundant than hydrogen.\\
The spectrum of silicon looks rather unexpected, exhibiting a knee-like
structure at around 3~PeV and decreasing very steeply above. Contrary to
silicon, the iron spectrum looks very flat in this representation with a
slight change of index to
$\gamma\approx-2.5$ above 10~PeV. This behaviour of the
heavy group spectra will be discussed in section \ref{sibdiscuss}.
\section{Discussion}\label{discuss_sec}
\subsection{All particle energy spectrum}\label{totspekdiscuss}
\begin{figure}[t]
\centering\epsfig{file=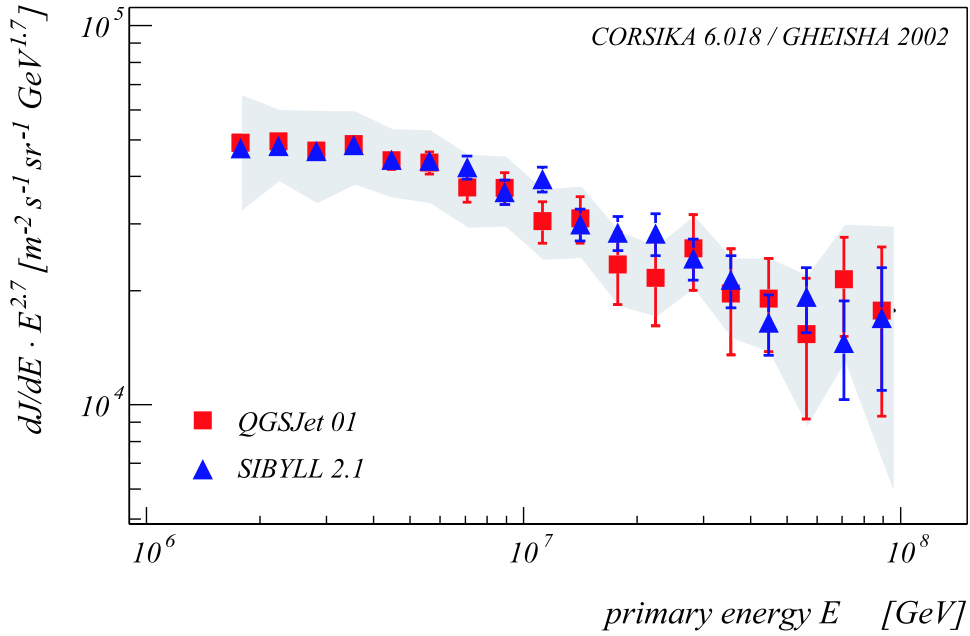,width=\linewidth}
\vspace{-1.3cm}
\caption{Result for the all particle energy spectrum using QGSJet and SIBYLL
simulations in the analysis. The shaded band represents the estimated
systematic uncertainties for the QGSJet solution which are of the same order
for the SIBYLL solution. For reasons of clarity only the QGSJet band is 
displayed.}
\label{totspek_pic}
\end{figure}

By summing up the five mass group spectra the all particle
spectrum is obtained. It is displayed in Fig.~\ref{totspek_pic} 
for both solutions.
The estimated statistical uncertainties are shown by the error bars, 
the shaded band represents the estimated systematic uncertainty, due
to the applied method (Gold algorithm) and the parameterization of
the tails of the shower size distribution, for the QGSJet results only. The
corresponding band for the SIBYLL solution is of same size and
omitted here for reasons of clarity. Tabulated values of the spectra are
given in Appendix \ref{flusswerte}.
\begin{figure}[b]
\centering\epsfig{file=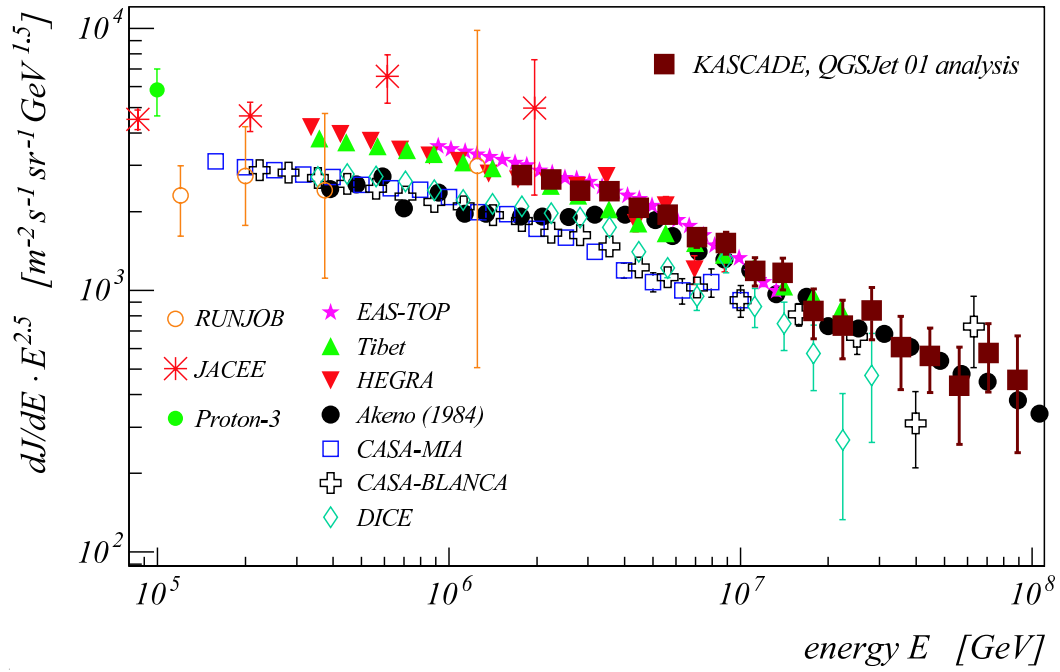,width=\linewidth}
\vspace{-0.6cm}
\caption{All particle spectrum for the QGSJet 01 based analysis in comparison
with results from RUNJOB \cite{apa01}, JACEE \cite{tak98}, Proton-3
\cite{gri70}, EAS-TOP \cite{agl99}, Tibet \cite{ame96}, HEGRA \cite{arq00},
Akeno \cite{nag84}, CASA-MIA \cite{gla99}, CASA-BLANCA \cite{fow01}, and
DICE \cite{swo00}.}
\label{allpart_comppic}
\end{figure}

The knee is clearly visible for both cases. The spectrum is fitted with the
expression~\cite{knieformel}
\begin{equation}
\frac{dJ(E)}{dE} = p_{0}\cdot
E^{p_{2}}\left(1+\left(\frac{E}{p_{1}}\right)^{p_{4}}
\right)^{(p_{3}-p_{2})/p_{4}},
\label{kneefunction}
\end{equation}
where $p_{1}$ corresponds to the knee position, $p_{2}$ and $p_{3}$ are
the spectral indices below and above the knee, and $p_{4}$ is a parameter
describing the sharpness of the knee.
In the case of the QGSJet 01 solution for the knee position
a value of $4.0\pm0.8$~PeV and for the spectral indices $-2.70\pm0.01$ and
$-3.10\pm0.07$ were obtained. For the SIBYLL solution the corresponding
values are $5.7 \pm 1.6$~PeV, 
$-2.70 \pm 0.06$, and $-3.14 \pm 0.06$. In both
cases, the fit is insensitive to the value of $p_{4}$ which was
therefore fixed to a value of 4.
The $\chi^{2}$ per degree of freedom is 0.35 in the QGSJet case, and 0.42 for
the SIBYLL solution.
Within statistical uncertainties the results for the two interaction models
coincide. It should be stressed that although the band of systematic
uncertainties might suggest the possibility of a spectrum 
without a knee, each of
the spectra defining this band exhibits a knee at around 5~PeV.

This result is essentially 
independent of the interaction models used and in good agreement with
results from other experiments. In Fig.~\ref{allpart_comppic} the
QGSJet based results are displayed together with results from some
other experiments.
Concerning the flux at the knee and the knee position
a very good agreement is especially reached with the HEGRA and
the EAS-TOP experiments.
\subsection{Description of data -- QGSJet based analysis}\label{qgsdiscuss}
\begin{figure}[b]
\centering\epsfig{file=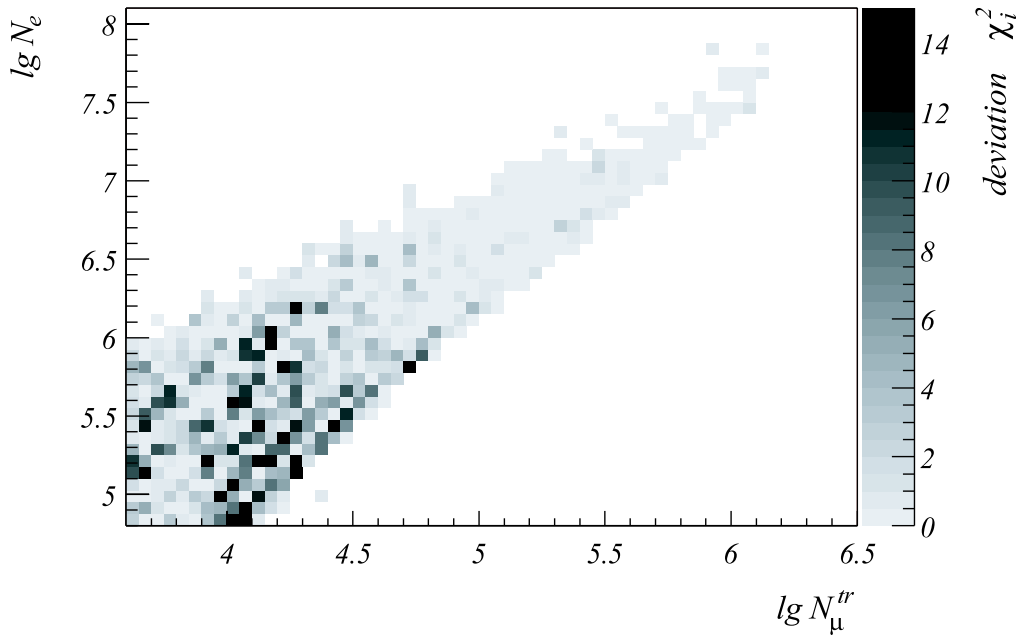,width=\linewidth}
\caption{Distribution of the individual $\chi^{2}_{i}$ in the data range for
the QGSJet solution.}
\label{qgs_chiplane}
\end{figure}

To judge on the properties and the quality of the solution
a vector
$\vec{Y}_{con}$ is ``constructed`` by forward folding of the solution
according to Eq.~(\ref{simplemateq}) and a $\chi^{2}$ test 
is performed. This results in a value of
$\chi^{2}_{dof}=2.38$.
The individual contributions $\chi^{2}_{i}$ of interval $i$ 
to this value are displayed in Fig.
\ref{qgs_chiplane} as a twodimensional distribution. 
It is obvious that the obtained solution is not
able to describe the data satisfactorily.
\begin{figure}[t]
\begin{minipage}[b]{.50\linewidth}
\centering\epsfig{file=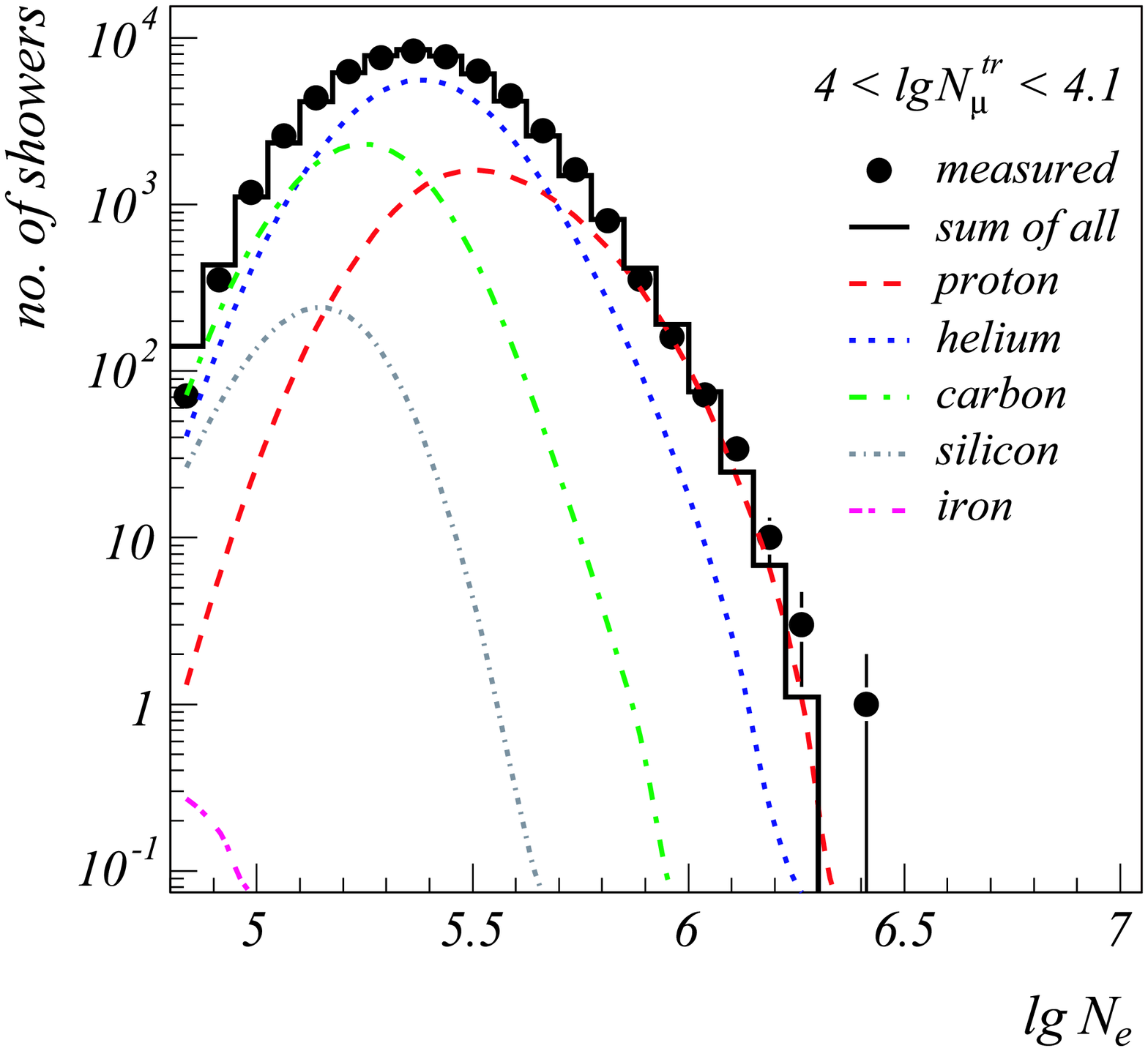,width=\linewidth}
\end{minipage}
\begin{minipage}[b]{.50\linewidth}
\centering\epsfig{file=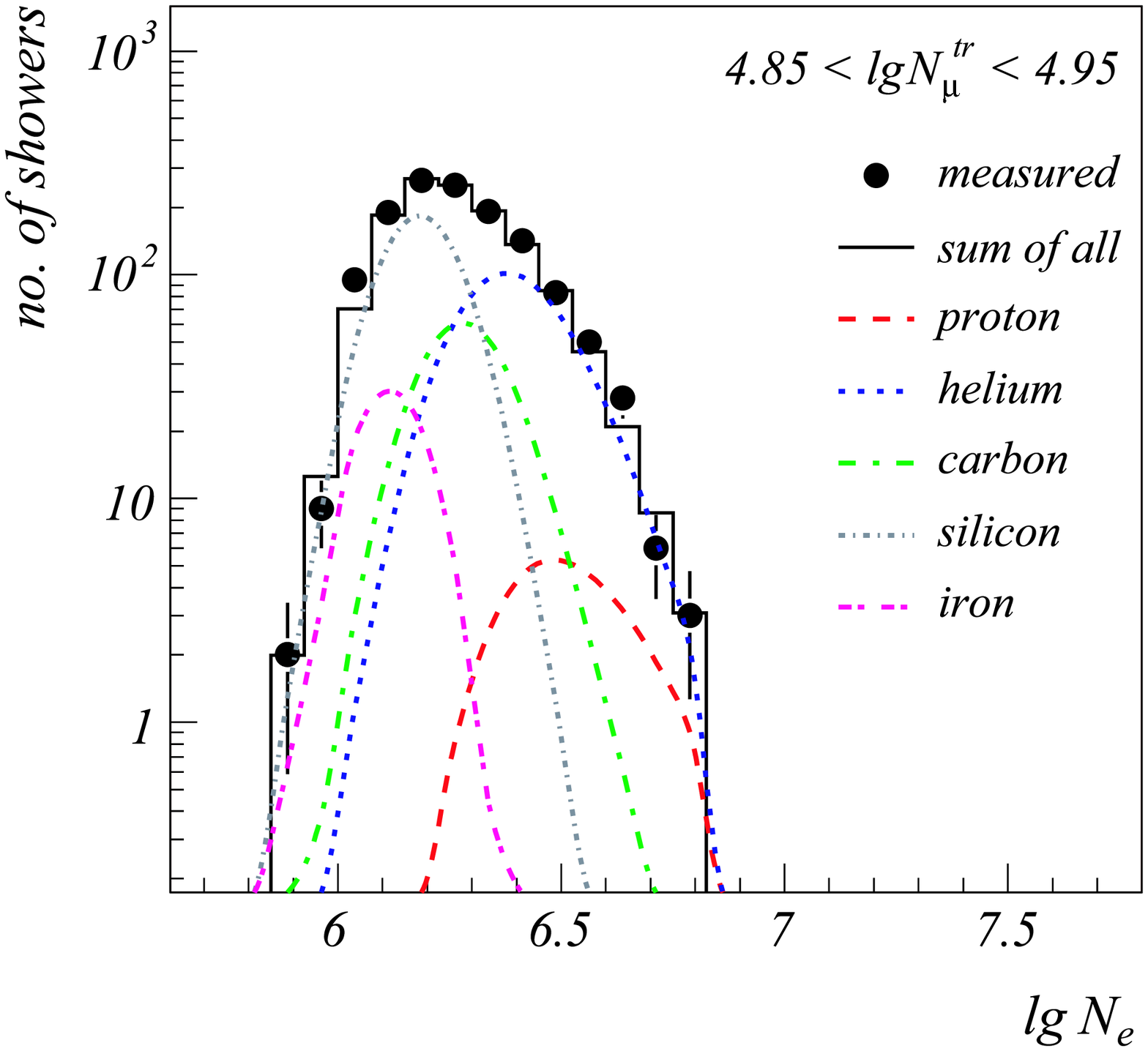,width=\linewidth}
\end{minipage}
\caption{Comparison of measured data with the QGSJet based solution. Left:
Electron size distribution for $4<\lg N_{\mu}^{tr}<4.1$. Right:
Electron size distribution for $4.85<\lg N_{\mu}^{tr}<4.95$.}
\label{QGSeledistmycut}
\end{figure}

The bulk of the deviations between measured and constructed data is
concentrated in the lower part of the measurement range, i.e. at low
energies, with a slight concentration in the region of small electron
numbers (for fixed muon number), i.e. showers induced by heavy primaries.
For higher energies (large shower sizes) the description of the data is
quite well within the statistical uncertainties.

To clarify the nature of these deviations, it is instructive to
look at the $\lg N_{e}$-distribution for given $\lg N_{\mu}^{tr}$ bins.
Fig.~\ref{QGSeledistmycut} displays the measured distributions (points)
for different $\lg N_{\mu}^{tr}$ bins together with 
the resulting distributions
of the forward folding (histogram). In addition, the
contribution of the different primary types are shown by smooth curves.
As can be seen, a large contribution of showers induced by light elements
is needed for small muon number bins to describe the
right tail of the distribution (large electron numbers).
As a consequence, no iron showers are needed for the
description of the left-hand tail of the distribution. Even with practically
no iron present at all there are still more showers with $\lg N_{e}<5$
calculated than measured. The situation improves 
for higher energies (large muon numbers). First, the description
of the distribution is better; second, now iron is also required to describe
the measured electron sizes. By investigating such figures the reason
for the negligible iron flux at low energies for the QGSJet result can be
understood.
Investigations of the $\lg N_{\mu}^{tr}$-distributions for different
$\lg N_{e}$ bins yield corresponding results.\\
To summarize, showers generated using QGSJet seem to predict too many muons or
too few electrons at low energies than required by the data. 
\clearpage
\subsection{Description of data -- SIBYLL based analysis}\label{sibdiscuss}
\begin{figure}[b]
\centering\epsfig{file=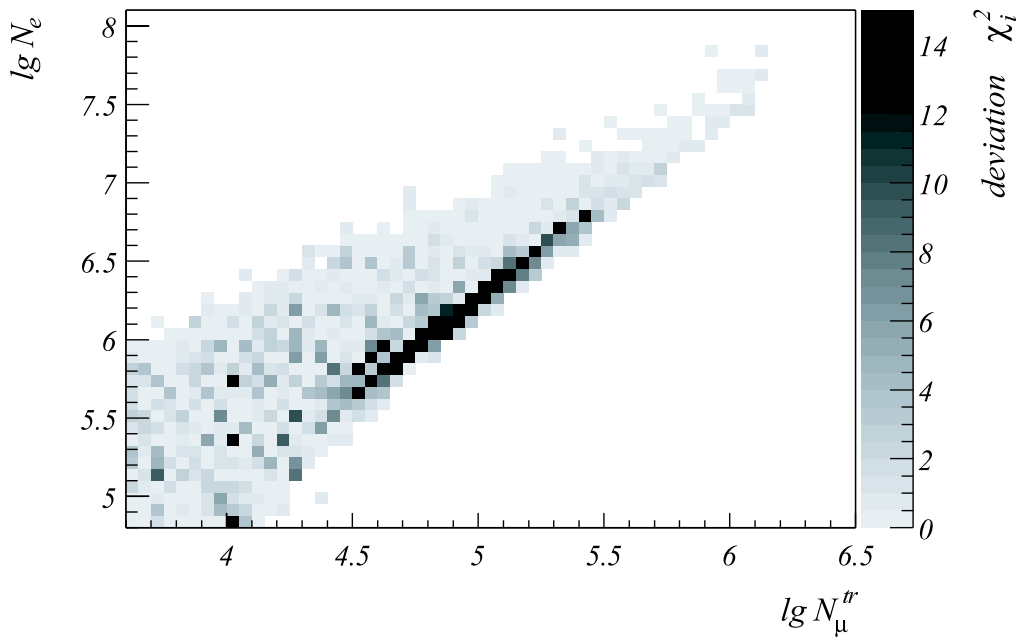,width=\linewidth}
\caption{Distribution of the individual $\chi^{2}_{i}$ in the data range for
the SIBYLL solution.}
\label{sib_chiplane}
\end{figure}

A similar comparison was performed using results based on the SIBYLL model.
An overall value for $\chi^{2}_{dof}$ of 2.46
was obtained,
being quite similar to the QGSJet case.
Again the solution
is not capable to describe the measured data in the complete region 
of measurement. 
The distribution of individual $\chi^{2}_{i}$, displayed in
Fig.~\ref{sib_chiplane}, is very 
different compared to the QGSJet based solution (Fig.~\ref{qgs_chiplane}).
The bulk of the deviations is concentrated at medium to high 
$\lg N_{\mu}^{tr}$ and small $\lg N_{e}$, i.e. in the region 
of heavy primaries with higher energies. 
Other than with the QGSJet solution, only small deviations occur at low
energies.
\begin{figure}[t]
\begin{minipage}[b]{.50\linewidth}
\centering\epsfig{file=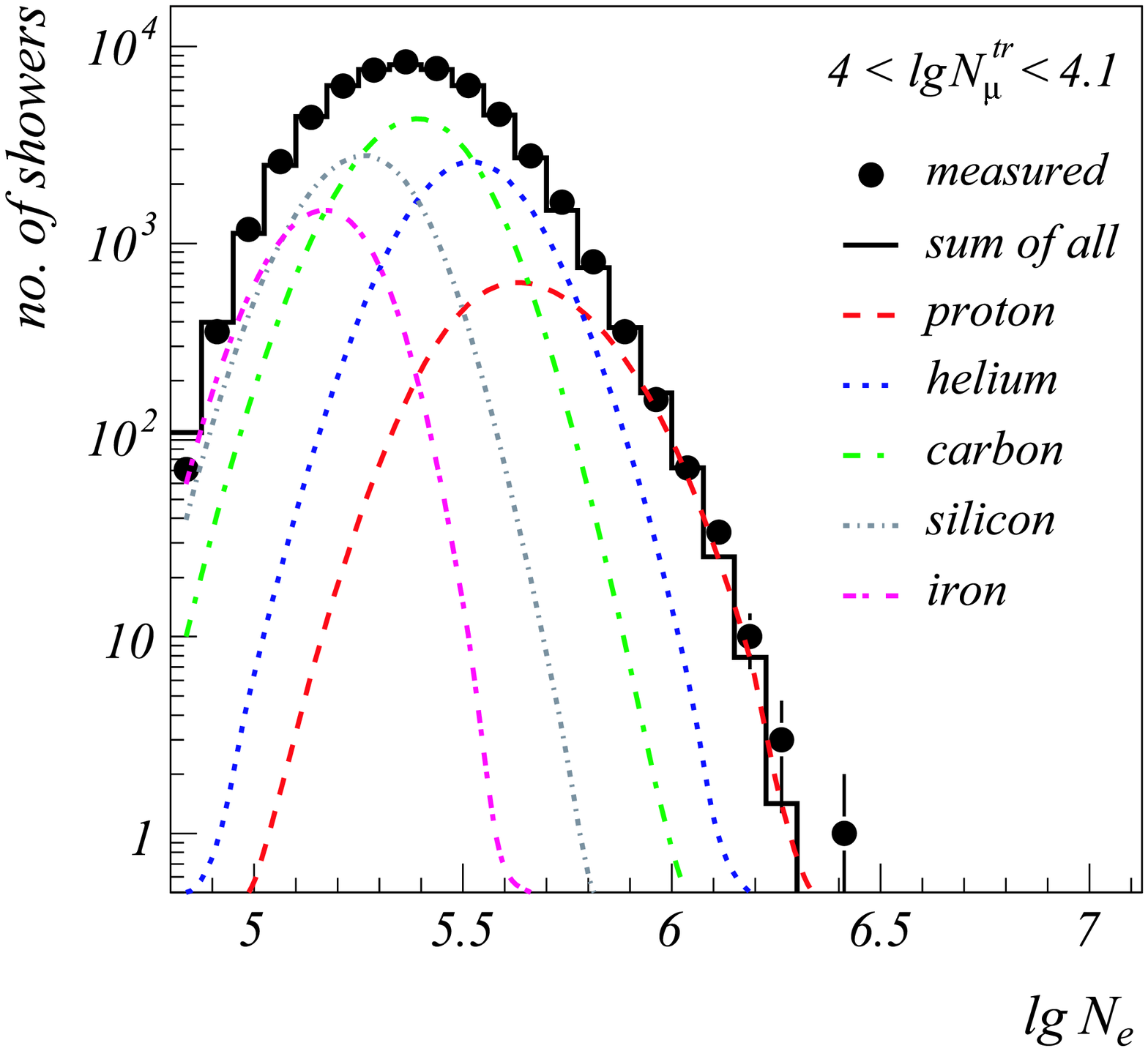,width=\linewidth}
\end{minipage}
\begin{minipage}[b]{.50\linewidth}
\centering\epsfig{file=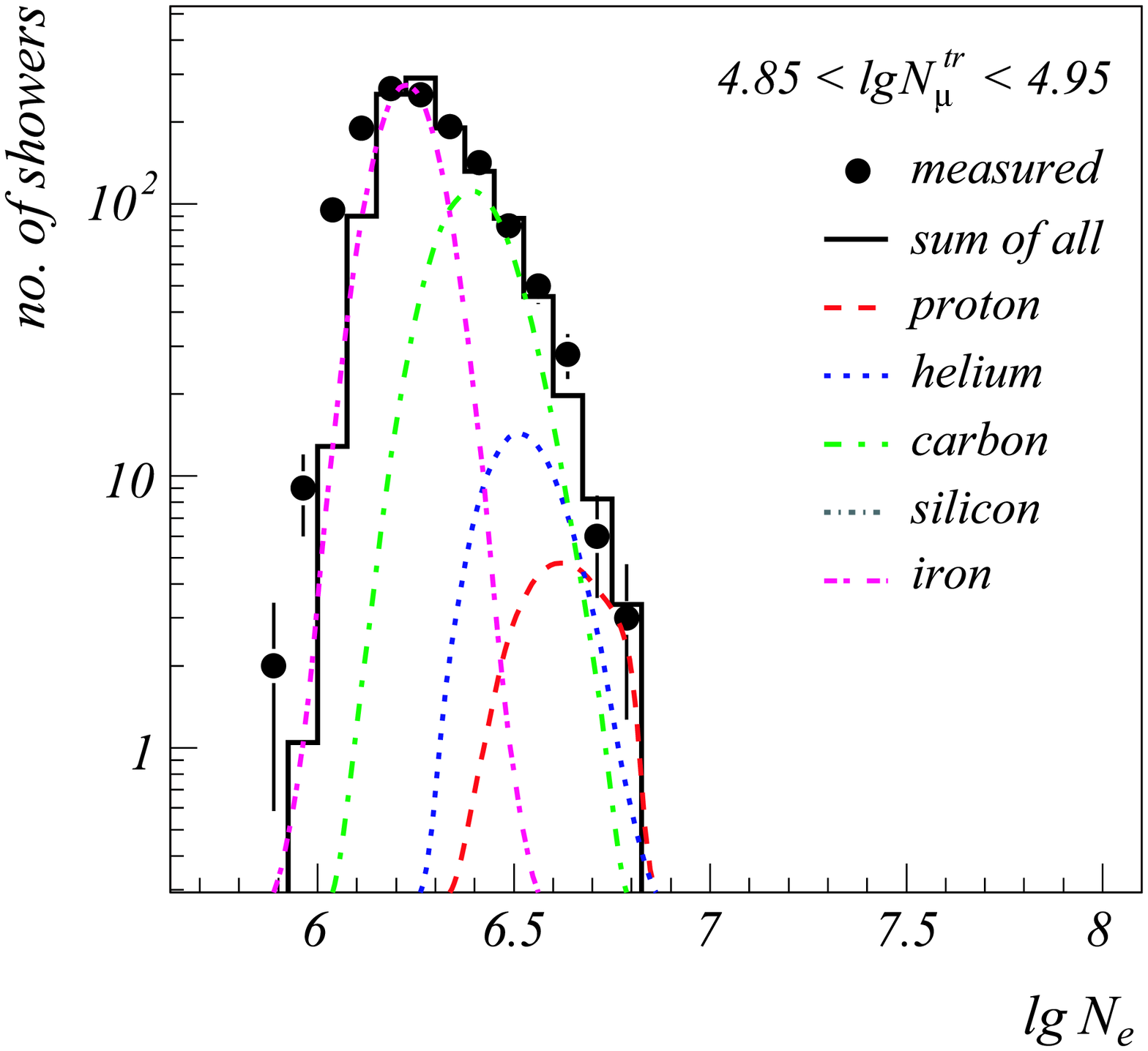,width=\linewidth}
\end{minipage}
\caption{Comparison of measured data with the SIBYLL based solution. Left:
Electron size distribution for $4<\lg N_{\mu}^{tr}<4.1$. Right:
Electron size distribution for $4.85<\lg N_{\mu}^{tr}<4.95$.}
\label{SIBeledistmycut}
\end{figure}

In Fig.~\ref{SIBeledistmycut} the measured electron size distributions
together with the constructed distributions are displayed for the same
$\lg N_{\mu}^{tr}$ bins as in Fig.~\ref{QGSeledistmycut}. The
description of the $\lg N_{e}$-distribution for low $\lg N_{\mu}^{tr}$
bins is much better than for the QGSJet based result, only small deviations
are found. Since the maximum of the $\lg N_{e}$-distribution for carbon
induced showers nearly coincides with that of the measured
distribution, a high abundances
of this mass group is found in case of the SIBYLL based solution.
The good description of the low energy data range by the SIBYLL based
simulations is also the reason for the smaller systematic uncertainties
in Fig.~\ref{SIB_unfoldbands} at lower energies when compared to the QGSJet
results. These uncertainties are dominated
by the unknown shape of the tails of the shower fluctuations $s_{A}$
and reflect the stability of the solution
against disturbances of the response matrix like changes of the distribution
tails. This stability is highest for well described data, resulting in smaller
uncertainties.

In contrast, the
description in the higher $\lg N_{\mu}^{tr}$ range is much worse than before.
This can be seen in the right part of Fig.~\ref{SIBeledistmycut}.
In particular the left-hand tail of the
$\lg N_{e}$-distribution cannot be described 
using the five assumed particle types. In order to fit
the distribution as well as possible the iron contribution has to be
raised nearly to the maximum value allowed by the maximum of the
observed distribution. On the other hand, the right tail towards higher
values of $\lg N_{e}$ can only be described using the lighter elements. As
a result, there is no space left for a significant contribution of silicon 
which explains
the sharp decrease in the silicon spectrum in Fig.~\ref{SIB_unfoldbands}.
Whereas the data description at lower energies works quite well, at higher
energies showers generated with the SIBYLL model seem to be too electron rich
or too muon poor compared to the data. The same conclusion 
holds when investigating the
$\lg N_{\mu}^{tr}$-distribution for different $\lg N_{e}$ bins.
\subsection{Some qualitative considerations -- open problems}
\begin{figure}[b]
\centering\epsfig{file=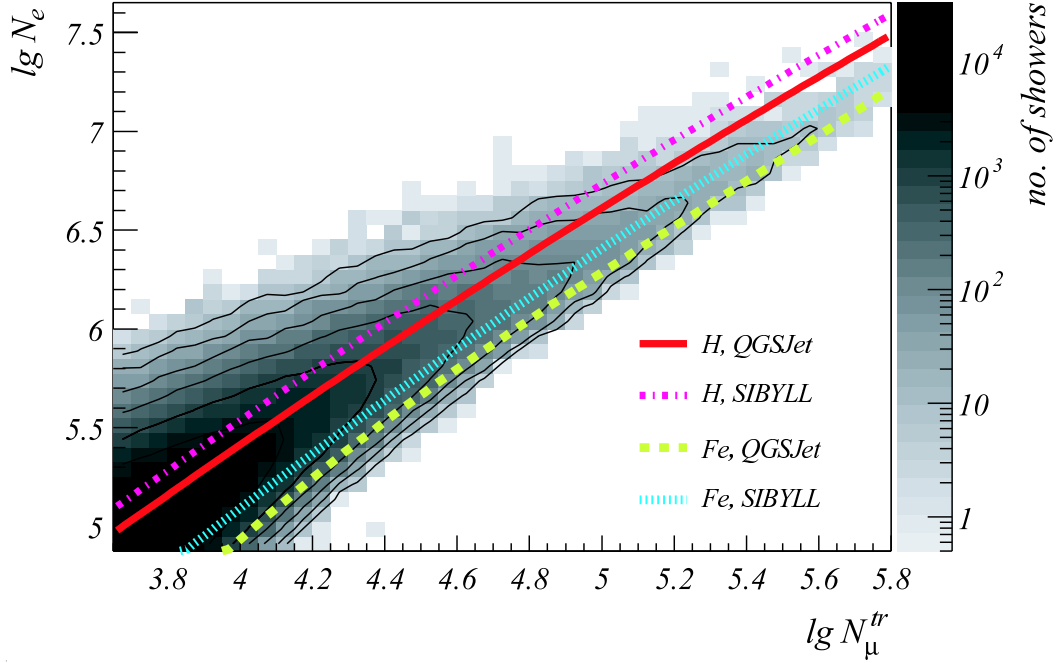,width=\linewidth}
\caption{Two-dimensional shower size spectrum of
$\lg N_{e}$ and $\lg N_{\mu}^{tr}$
together with isolines and lines of the most probable values for proton and
iron induced showers for both simulations.}
\label{data_iso_mop}
\end{figure}

Despite the fact that none of the two hadronic interaction models is able
to describe the whole data range consistently, it is possible to get hints,
why their predictions do not match the data and how
agreement with the data could be improved.
In Fig.~\ref{data_iso_mop} a part of the two-dimensional size spectrum
and in addition some lines of constant intensity
({\it isolines}) are drawn. Overlaid are lines representing the
energy dependence of the maximum value of the probability distributions
$p_{A}(\lg N_{e}, \lg N_{\mu}^{tr}|\lg E)$ of Eq.~(\ref{integral1}), i.e. the
most probable pair of the shower size values. These lines of the most
probable values are displayed for the primary particles hydrogen (proton)
and iron and for the two simulation sets.

It is noticeable that the lines of the most probable values for the two
interaction models are nearly parallel. For SIBYLL simulations these lines
are evenly shifted towards larger values of $\lg N_{e}$ and smaller values of
$\lg N_{\mu}^{tr}$ with respect to the corresponding lines for QGSJet
simulations.

For the SIBYLL based simulations the line of the most probable values of iron
showers tends to lie the more in the central region of the data the 
higher the energy and moves away from the lower edge (small
$\lg N_{e}$ for fixed $\lg N_{\mu}^{tr}$) of the data distribution. This
lower edge is expected to be dominated by showers induced by heavy elements.
Consistent with this relatively large distance to the lower edge are the
discrepancies in the description of the measured data by the SIBYLL based
results for higher energies (Fig.~\ref{SIBeledistmycut}). One way to solve
this problem would be a reduction of the predicted electron number in
high energy SIBYLL simulations which would result in a decreasing slope of
the corresponding lines of the most probable values. Another possibility
might be a weaker decrease of shower fluctuations in SIBYLL simulations with
increasing energy.

In the case of the QGSJet based solution the description of the data at
higher energies is better whereas discrepancies occur at lower energies
(see Fig.~\ref{QGSeledistmycut}). In the region of small shower sizes no
contribution of the heavy component is needed to describe the data.
Referring to Fig.~\ref{data_iso_mop}, one approach would be to shift
the lines of the most probable values in the region of smaller shower sizes
away from the lower edge
of the data distribution. This would mean that more electrons or
fewer muons are predicted for showers at low energies. Another possibility
could be the reduction of shower fluctuations at lower energies
for QGSJet based simulations.

One might argue that the low energy hadronic interaction model could 
influence the simulations and contribute to the differences. 
However, when using the FLUKA~\cite{fluka} instead
of the GHEISHA code we found almost no changes for
the electron and muon shower size distributions. Therefore, FLUKA is not able
to improve the situation significantly.

Although these considerations are qualitative they may give some hints for
further improvement of hadronic interaction models.
Investigations with a toy model which consists of simply shifting
the predictions of SIBYLL based simulations towards QGSJet based predictions
with increasing energy,
resulted in a consistent description of the measured data. (We are
aware that this is not a very reasonable procedure.)
\subsection{Systematic model uncertainties - the proton spectrum}
\begin{figure}[t]
\centering\epsfig{file=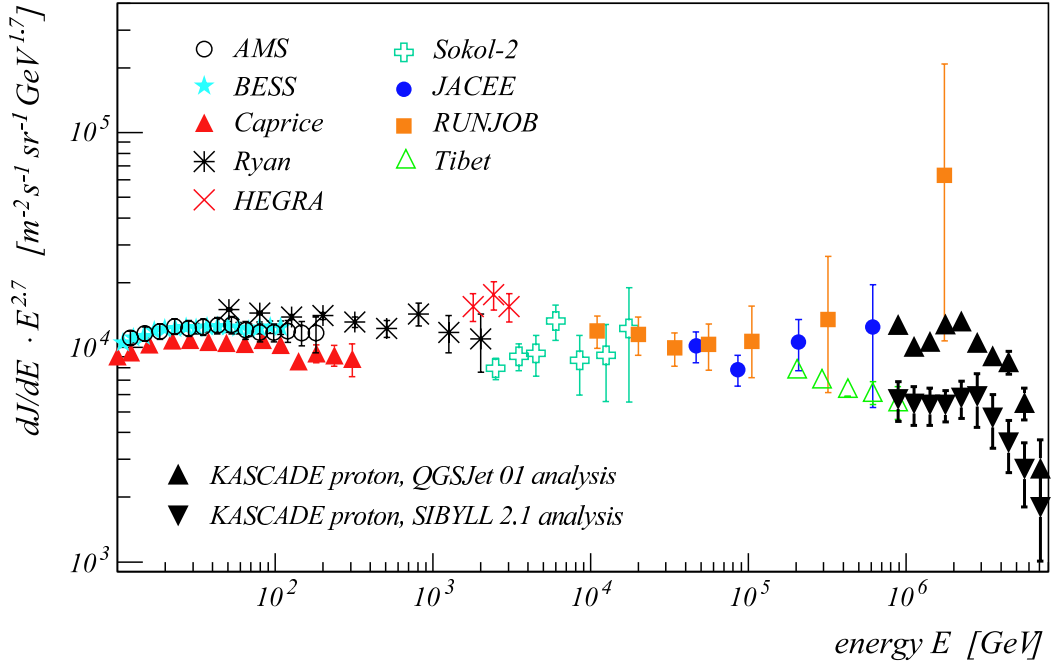,width=\linewidth}
\caption{Results for the proton energy spectrum for both of 
our analysis together with results from direct (AMS\cite{agu02},
BESS\cite{san00}, CAPRICE\cite{boe03}, Ryan\cite{rya72}, SOKOL-2
\cite{gri90}, RUNJOB\cite{apa01}, JACEE\cite{tak98}) and indirect
(HEGRA\cite{aha99}, Tibet\cite{ame00}) measurements.}
\label{prot_connect}
\end{figure}

Since neither of the interaction models used in the analysis can describe
the measured data consistently the results for the individual element
spectra are not fully reliable. However, the difference between 
the results for the spectrum
of one particular particle can serve as an estimate of the systematic
uncertainty due to the hadronic interaction models.

To visualize this uncertainty, it is instructive to compare
direct and indirect measurements of protons at lower energies
with the corresponding spectra of our analyses (Fig.~\ref{prot_connect}).
Due to the low abundances of elements lighter than carbon but other than
proton and helium the results for proton would give the real
spectrum of the single element in the case of correct simulations. Since the
elements carbon,
silicon, and iron stand for elemental groups, which are
loosely defined, a comparison with data from direct measurements is not
possible for these heavier elements.

Despite the large difference between our two results
they are in good agreement with the extrapolations of those of
balloon-borne experiments for the proton spectrum.
At present, the statistical uncertainties of direct measurements above
$10^{14}$ eV are of the same order of magnitude as the systematic
uncertainty of air shower based analyses due to the hadronic interaction
models. Further improvement requires a more reliable 
theoretical description of high energy hadronic interactions.
\section{Summary and conclusion}\label{conclus_sec}
Using the two-dimensional shower size spectrum of electron number
$\lg N_{e}$ and
muon number $\lg N_{\mu}^{tr}$ measured with KASCADE an analysis was
presented yielding energy spectra for five primary mass groups,
representing the
chemical composition of cosmic rays. For this analysis, air shower
simulations with two different high energy hadronic interaction models
(QGSJet 01 and SIBYLL 2.1) were used.
The reconstructed all particle spectra for
both simulation sets
coincide within the statistical and systematic uncertainties and are
consistent with results from other experiments. The knee is observed at an
energy around $\approx5$~PeV with a change of index $\Delta\gamma\approx 0.4$.
The situation differs quite strongly when considering the results of the
mass group
spectra. Common is the appearence of knee-like features in the
spectra of the light elements.
For both models the position of the knees in these spectra is shifted
towards higher energy with increasing element number.
A closer inspection revealed that none of the two interaction models is
capable of describing the measured data consistently over the whole
measurement range. For the QGSJet based analysis deviations occur at low
energies whereas for the SIBYLL based analysis the higher energies are
problematic.

Summarizing, it has been demonstrated that unfolding 
methods are capable to reconstruct
energy spectra of individual mass groups from air shower data, in
addition to the all particle spectrum.
At present, the limiting factors of the analysis are the properties
of the high energy interaction models used and not the
quality or the understanding of the KASCADE data. The observed
discrepancies between simulations and data have to be attributed 
to the models and may give
valuable information for their further improvements.
\begin{ack}
The authors would like to thank the members of the engineering and technical
staff of the KASCADE collaboration who contributed with enthusiasm and
commitment to the success of the experiment. The KASCADE experiment is
supported by the German Federal Ministry of Education and Research and was
embedded in collaborative WTZ projects between Germany and Romania (RUM
97/014), Poland (POL 99/005), and Armenia (ARM 98/002). The Polish group
acknowledges support by KBN research grant 1 P03B 03926 for the years
2004-2006.
\end{ack}
\begin{appendix}
\section{The matrix equation and unfolding methods}
\subsection{Formulation of the problem as matrix equation}
\label{mathe}
The bin content $N_{i}$ of each cell of the two-dimensional shower size
spectrum displayed in Fig.~\ref{useddata_pic} can be written as
\begin{equation} \hspace{-8mm}
N_{i} = 2\pi A_{s}T_{m}\sum_{A=1}^{N_{A}}
\int_{0^{\circ}}^{18^{\circ}}
\int_{-\infty}^{+\infty} \frac{dJ_{A}}{d\lg E} \/ p_{A}((\lg N_{e},
\lg N_{\mu}^{tr})_{i} | \lg E) \frac{1}{2}\sin 2\theta\,d\lg E\,d\theta
\label{integral_A1}
\end{equation}
with the differential flux $dJ_{A}/d\lg E$ of an
element of mass number $A$ and the conditional probability
$p_{A}$ describing the probability to measure a shower of primary energy
$\lg E$ and primary mass $A$ with shower sizes $(\lg N_{e},
\lg N_{\mu}^{tr})_{i}$. For sufficiently large showers, which are 
only included in the present analysis,
measurement time $T_{m}$ and sampling area $A_{s}$
can be treated as constants and no dependence on
azimuth angle is present, resulting in the factor $2\pi$.

The probability $p_{A}$ can be expressed as
\begin{equation}
p_{A} = \int_{-\infty}^{+\infty}
\int_{-\infty}^{+\infty}
s_{A} \epsilon_{A} r_{A} \,d\lg N_{e}^{true} \,d\lg N_{\mu}^{tr,true}
\label{integral_A2}
\end{equation}
with the primary dependent intrinsic shower fluctuations
$s_{A}$, the
properties of the reconstruction (resolution and systematic shifts)
$r_{A}$ and the combined efficiencies for
detection and reconstruction
$\epsilon_{A}$.

Simulation studies have shown that at KASCADE efficiencies
$\epsilon_{A}$ and reconstruction properties $r_{A}$ do not
depend on zenith angle $\theta$ for $\theta < 20^{\circ}$.
In addition the angular
resolution of the KASCADE array in the considered shower size range is better
than 0.2$^{\circ}$, so effects due to limited angular resolution
can be savely neglected. Since only showers with
$0^{\circ} \leq \theta < 18^{\circ}$ are considered,
the integration over zenith angle
can be incorporated into $s_{A}$. In this sense Eq. \ref{integral_A1}
can be
written as 
\begin{equation}
N_{i} = A_{s}T_{m} \Delta\Omega \sum_{A=1}^{N_{A}}
\int_{-\infty}^{+\infty} \frac{dJ_{A}}{d\lg E} \/ p_{A}((\lg N_{e},
\lg N_{\mu}^{tr})_{i} | \lg E) \,d\lg E
\label{integral_A3}
\end{equation}
with effective solid angle $\Delta\Omega$. The mentioned integration over
zenith angle is now included in the shower fluctuations.

Using the abbreviation
\begin{equation}
x_{j}^{A} = A_{s}T_{m}\Delta\Omega
\int_{\lg E_{j}}^{\lg E_{j}+\Delta\lg E} \frac{dJ_{A}}{d\lg E} \,d\lg E
\end{equation}
the integral can be written as a sum over $n$ energy intervals of width
$\Delta\lg E$ with $\lg E_{j}$ denoting the
lower bin edges:
\begin{equation}
N_{i} = \sum_{A=1}^{N_{A}} \sum_{j=1}^{n} R_{ij}^{A} \/ x_{j}^{A}
\end{equation}
Here the matrix element $R_{ij}^{A}$ is defined by
\begin{equation}
R_{ij}^{A} = \displaystyle \frac{\displaystyle
\int_{\lg E_{j}}^{\lg E_{j}+\Delta\lg E} \frac{dJ_{A}}{d\lg E} \/
p_{A}((\lg N_{e},\lg N_{\mu}^{tr})_{i} | \lg E) \,d\lg E}
{\displaystyle
\int_{\lg E_{j}}^{\lg E_{j}+\Delta\lg E} \frac{dJ_{A}}{d\lg E} \,d\lg E}.
\label{matrixelement}
\end{equation}
For small bin width $\Delta\lg E$ the value of the matrix elements
$R_{ij}^{A}$ are not sensitive to the correct shape
of the differential energy spectra $dJ_{A}/d\lg E$.
A decoupling between the matrix element $R_{ij}^{A}$ and the
fluxes $dJ_{A}/d\lg E$ is then achieved. In the present analysis a bin
width of $\Delta\lg E = 0.1$ is chosen which turns out 
to be sufficiently small.

Introducing the $m$-dimensional data vector $\vec{Y}$ which contains the $m$ 
cell contents $N_{i}$ of the two-dimensional shower size spectrum, the
relation between data and energy spectra can be written as
\begin{equation}
\vec{Y} = \sum_{A=1}^{N_{A}} {\bf R^{A}} \vec{X^{A}} \quad{\mathrm{with}}\quad
\vec{X^{A}} = \left( \begin{array}{l}
x_{1}^{A}\\ x_{2}^{A}\\ \vdots 
\end{array} \right)
\quad{\mathrm{and}}\quad
\vec{Y} = \left( \begin{array}{l}
N_{1}\\ N_{2}\\ \vdots 
\end{array} \right)
\label{matrixsumme}
\end{equation}
with the elements of the matrix ${\bf R^{A}}$ defined by
Eq.~(\ref{matrixelement}).
For a more compact notation the summation over different primaries
can be incorporated into the matrix equation by defining the response
matrix ${\bf R}$ and the vector of unknowns $\vec{X}$ schematically through
\begin{equation}
{\mathbf R} = \left({\mathbf R}^{1}\quad{\mathbf R}^{2}\quad
{\ldots} \right)
\quad{\mathrm{and}}\quad \vec{X} = \left( \begin{array}{l}
\vec{X}^{1}\\ \vec{X}^{2}\\ \vdots 
\end{array} \right),
\end{equation}
where the response matrix ${\bf R}$ is a block matrix consisting of the
response matrices ${\bf R^{A}}$ for the individual particles. Adopting
this notation yields for Eq.~(\ref{matrixsumme}) the simple expression
\begin{equation}
\vec{Y} = {\bf R} \vec{X}.
\label{simple_append}
\end{equation}

\subsection{Unfolding methods used}
\label{unfold_app}
\subsubsection{The Gold algorithm}
For the application of the Gold algorithm \cite{gold} 
a slight modification of
Eq.~(\ref{simple_append}) is necessary.
A new data vector $\vec{Y}_{mod}$ and a new
response matrix ${\bf \tilde{R}}$ are defined 
via the diagonal matrix ${\bf C}$ containing 
the statistical uncertainties of the data:
\begin{equation}
{\bf \tilde{R} = R^{T}CCR}\quad \mathrm{and} \quad
\vec{Y}_{mod} = {\bf R^{T}CC}\vec{Y}, \quad \mathrm{yielding} \quad
{\bf \tilde{R}}\vec{X} = \vec{Y}_{mod}.
\label{mateqmod}
\end{equation}
In case of existence of a solution of Eq.~(\ref{mateqmod})
the Gold algorithm constructs iteratively the diagonal
matrix ${\bf D}$
with elements $d_{ii} = x_{i}/y_{mod,i}$ which yields the desired
vector $\vec{X}$ simply as $\vec{X} = {\bf D}\vec{Y}_{mod}$. The
iterative prescription for the components reads
\begin{equation}
x_{i}^{k+1} = \frac{x_{i}^{k} y_{mod,i}}{\displaystyle \sum_{j=1}^{n}
\tilde{R}_{ij}x_{j}^{k}}
\end{equation}
where $x_{i}^{k}$ is the estimated solution in the $k^{\mathrm{th}}$
iteration step.
\subsubsection{Bayesian unfolding}
The unfolding procedure based on the Bayesian theorem \cite{bayes}
constructs, like the Gold algorithm, iteratively
a matrix ${\bf P}$. The elements of ${\bf P}$ contain the
probabilities for the values $x_{i}$ if the data $\vec{Y}$ is measured.
Here, ${\bf P}$ is not a diagonal matrix.
The unknown vector $\vec{X}$ is calculated by $\vec{X} = {\bf P}\vec{Y}$.
Since ${\bf P}$ is not a square matrix, one can directly start with
Eq.~(\ref{simple_append}). The iterative prescription for the components
$x_{i}$ reads
\begin{equation}
x_{i}^{k+1} = \frac{1}{\sum_{j=1}^{m}R_{ji}}
\displaystyle\sum_{j=1}^{m} \frac{R_{ji}x_{i}^{k}}
{\sum_{l=1}^{n}R_{jl}x_{l}^{k}}y_{j}
\end{equation}
with $x_{i}^{k}$ being the estimated solution after $k$ iteration steps.
\subsubsection{Entropy based unfolding}
The entropy based method \cite{schmelling} is a special case of
regularized unfolding.
The basic idea is the minimization of an extented $\chi^{2}$-functional
with the incorporation of some constraints.
The modified
$\chi^{2}_{mod}$-functional reads 
\begin{equation} \hspace{-5mm}
\chi^{2}_{mod} = \sum_{j=1}^{m}\displaystyle\frac{(y_{j}-\sum_{i=1}^{n}
R_{ji}x_{i})^{2}}{\sigma(y_{j})^{2}} + \tau S(\vec{X})
\quad\mathrm{with}\quad S(\vec{X}) = \sum_{i=1}^{n}x_{i}\ln\frac{x_{i}}{r_{i}}
\label{redEnt}
\end{equation}
with $\sigma(y_{j})$ being the statistical uncertainty of the data element
$y_{j}$.
$S(\vec{X})$ is the entropy-based functional depending on the solution vector
$\vec{X}$, $\tau$ the so-called {\it regularization parameter} governing the
influence of the regularization term $S$. The values $r_{i}$ are the elements
of a reference distribution vector $\vec{r}$ which can be considered as the
best guess of the solution $\vec{X}$. Depending on the value of the
regularization parameter $\tau$ the solution $\vec{X}$ resembles more or less
the reference $\vec{r}$. In this way a balance between statistical
significance of the solution and systematic uncertainty due to $\vec{r}$ is
achieved. For the minimization of Eq.~(\ref{redEnt}) the MINUIT
\cite{minuit} package was used.
\newpage
\section{Tabulated values of the all particle spectrum}\label{flusswerte}
\begin{table}[h]
\caption{Differential flux values of the all particle energy spectrum for
QGSJet 01 and SIBYLL 2.1 based analysis. The first column of errors denotes
the statistical uncertainty, the second column the systematic uncertainty.}
\label{fluxtable}
\vspace{4mm}
\begin{tabular}{lll}
\hline
energy&d$J$/d$E$ $\pm$ stat. $\pm$ syst. (QGSJet)&d$J$/d$E$ $\pm$
stat. $\pm$ syst. (SIBYLL) \\
$[$GeV$]$ & [m$^{-2}$ s$^{-1}$ sr$^{-1}$ GeV$^{-1}$] &
[m$^{-2}$ s$^{-1}$ sr$^{-1}$ GeV$^{-1}$] \\
\hline
1.78$\cdot10^{6}$ &$ (6.54\pm0.25\pm2.20)\cdot10^{-13}$&
$(6.33\pm0.21\pm1.31)\cdot10^{-13}$ \\
2.24$\cdot10^{6}$ &$ (3.54\pm0.13\pm0.75)\cdot10^{-13}$&
$(3.45\pm0.14\pm0.70)\cdot10^{-13}$ \\
2.82$\cdot10^{6}$ &$ (1.80\pm0.08\pm0.49)\cdot10^{-13}$&
$(1.80\pm0.09\pm0.38)\cdot10^{-13}$ \\
3.55$\cdot10^{6}$ &$ (1.01\pm0.05\pm0.22)\cdot10^{-13}$&
$(1.00\pm0.05\pm0.22)\cdot10^{-13}$ \\
4.47$\cdot10^{6}$ &$ (4.90\pm0.27\pm1.00)\cdot10^{-14}$&
$(4.91\pm0.27\pm1.02)\cdot10^{-14}$ \\
5.62$\cdot10^{6}$ &$ (2.59\pm0.18\pm0.56)\cdot10^{-14}$&
$(2.62\pm0.14\pm0.55)\cdot10^{-14}$ \\
7.08$\cdot10^{6}$ &$ (1.20\pm0.11\pm0.26)\cdot10^{-14}$&
$(1.36\pm0.10\pm0.28)\cdot10^{-14}$ \\
8.91$\cdot10^{6}$ &$ (6.41\pm0.62\pm1.35)\cdot10^{-15}$&
$(6.26\pm0.46\pm1.30)\cdot10^{-15}$ \\
1.12$\cdot10^{7}$ &$ (2.81\pm0.35\pm0.59)\cdot10^{-15}$&
$(3.63\pm0.28\pm0.75)\cdot10^{-15}$ \\
1.41$\cdot10^{7}$ &$ (1.54\pm0.22\pm0.33)\cdot10^{-15}$&
$(1.48\pm0.14\pm0.31)\cdot10^{-15}$ \\
1.78$\cdot10^{7}$ &$ (6.24\pm1.35\pm1.39)\cdot10^{-16}$&
$(7.57\pm0.78\pm0.16)\cdot10^{-16}$ \\
2.24$\cdot10^{7}$ &$ (3.09\pm0.78\pm0.64)\cdot10^{-16}$&
$(4.05\pm0.51\pm0.87)\cdot10^{-16}$ \\
2.82$\cdot10^{7}$ &$ (1.98\pm0.45\pm0.43)\cdot10^{-16}$&
$(1.87\pm0.23\pm0.44)\cdot10^{-16}$ \\
3.55$\cdot10^{7}$ &$ (8.10\pm2.52\pm1.93)\cdot10^{-17}$&
$(8.81\pm0.14\pm2.38)\cdot10^{-17}$ \\
4.47$\cdot10^{7}$ &$ (4.22\pm1.16\pm1.14)\cdot10^{-17}$&
$(3.65\pm0.66\pm1.18)\cdot10^{-17}$ \\
5.62$\cdot10^{7}$ &$ (1.83\pm0.74\pm0.79)\cdot10^{-17}$&
$(2.29\pm0.45\pm0.89)\cdot10^{-17}$ \\
7.08$\cdot10^{7}$ &$ (1.37\pm0.40\pm0.53)\cdot10^{-17}$&
$(9.29\pm2.72\pm5.38)\cdot10^{-18}$ \\
8.91$\cdot10^{7}$ &$ (6.07\pm2.87\pm4.02)\cdot10^{-18}$&
$(5.81\pm2.07\pm4.31)\cdot10^{-18}$ \\
\hline
\end{tabular}
\end{table}
\clearpage
\end{appendix}

\end{document}